\documentclass{aa}  
\usepackage{courier,graphicx,epsfig,color,colordvi,natbib,url,twoopt,ulem,multirow}
\usepackage[varg]{txfonts}
\pdfoutput=1
\usepackage[breaklinks=true, draft]{hyperref} 
\usepackage{pdfcomment,acronym}

\bibpunct{(}{)}{;}{a}{}{,}    

\makeatletter
\newcommand{\bibnote}[2]{\@namedef{#1note}{#2}}
\newcommand{\biblink}[2]{\@namedef{#1link}{#2}}
\makeatother

\makeatletter
 \newcommandtwoopt{\citeads}[3][][]{%
   \nonstopmode
   \href{http://adsabs.harvard.edu/abs/#3}%
        {\def\hyper@linkstart##1##2{}%
         \let\hyper@linkend\@empty\citealp[#1][#2]{#3}}
   \biblink{#3}{\href{http://adsabs.harvard.edu/abs/#3}{ADS}}%
   \errorstopmode}            
 \newcommandtwoopt{\citepads}[3][][]{%
   \nonstopmode
   \href{http://adsabs.harvard.edu/abs/#3}%
        {\def\hyper@linkstart##1##2{}%
         \let\hyper@linkend\@empty\citep[#1][#2]{#3}}
   \biblink{#3}{\href{http://adsabs.harvard.edu/abs/#3}{ADS}}%
   \errorstopmode}            
 \newcommandtwoopt{\citetads}[3][][]{%
   \nonstopmode
   \href{http://adsabs.harvard.edu/abs/#3}%
        {\def\hyper@linkstart##1##2{}%
         \let\hyper@linkend\@empty\citet[#1][#2]{#3}}
   \biblink{#3}{\href{http://adsabs.harvard.edu/abs/#3}{ADS}}%
   \errorstopmode}            
 \newcommandtwoopt{\citeyearads}[3][][]{%
   \nonstopmode
   \href{http://adsabs.harvard.edu/abs/#3}%
        {\def\hyper@linkstart##1##2{}%
         \let\hyper@linkend\@empty\citeyear[#1][#2]{#3}}
   \biblink{#3}{\href{http://adsabs.harvard.edu/abs/#3}{ADS}}%
   \errorstopmode}            
\makeatother

\def\specchar#1{{\sc{#1}}}    

\def\vlos{\hbox{$v_{\rm{los}}$}}
\def\vturb{\hbox{$v_{\rm{micro}}$}}
\def\Blos{\hbox{$B_{\rm{los}}$}}
\def\Blosnlte{\hbox{$B_{\rm{los, NLTE}}$}}
\def\Bloslte{\hbox{$B_{\rm{los, LTE}}$}}
\def\BlosWFA{\hbox{$B_{\rm{los, WFA}}$}}
\def\BlosSTiC{\hbox{$B_{\rm{los, STiC}}$}}
\def\Bhor{\hbox{$B_{\rm{hor}}$}}

\def\Btot{\hbox{$B_{\rm{tot}}$}}

\def\ltau{\hbox{log $\tau_{500}$}}
\def\dltau{\hbox{$\Delta$log $\tau_{500}$}}

\def\SST{Swedish 1-m Solar Telescope}

\def\Hinode{{\it Hinode\/}}

\def\HMI{{\it Helioseismic and Magnetic Imager\/}}
\def\SDO{{\it Solar Dynamics Observatory\/}}

\def\stic{STockholm Inversion Code}
\def\ie{i.e.}
\def\eg{e.g.}

\def\CaIR{\mbox{Ca\,\specchar{ii}\,\,8542\,\AA}} 
\def\CaII{\mbox{Ca\,\specchar{ii}}}

\def\deg{\hbox{$^\circ$}}       

\def\HeI{\mbox{He\,\specchar{i}}} 
\def\HeIDthree{\mbox{He\,\specchar{i}\,\,D$_{3}$}}
\def\FeI{\mbox{Fe\,\specchar{i}}} 
\def\Halpha{\mbox{H\hspace{0.1ex}$\alpha$}} 

\def\SiI{\mbox{Si\,\specchar{i}}} 
\def\chisq{\mbox{$\chi^{2}$}}
\newcommand{\fov}[2]{{{#1}\arcsec$\times${#2}\arcsec}}

\newcommand{\XYis}[2]{\mbox{$(X,Y)=({#1}\arcsec,{#2}\arcsec)$}}

\newcommand{\xyisMm}[2]{\mbox{$(x,y)=({#1}\,{\rm{Mm}},{#2}\,{\rm{Mm}})$}}

\newcommand{\rev}[1]{{{#1}}}

\def\figspath{./}

\begin{document}

\title{Non-LTE inversions of a confined X2.2 flare: \\ 
I. Vector magnetic field in the photosphere and chromosphere
}
\titlerunning{Non-LTE inversions of a confined X2.2 flare: I. Vector magnetic
field}
\authorrunning{G.~J.~M.~Vissers et.~al.} 
 
\author{G.~J.~M.~Vissers$^{1}$ \and 
S.~Danilovic$^{1}$ \and
J.~de la Cruz Rodr{\'i}guez$^{1}$ \and
J.~Leenaarts$^{1}$ \and
R.~Morosin$^{1}$ \and
C.~J.~D{\'i}az Baso$^{1}$ \and
A.~Reid$^{2}$ \and
J.~Pomoell$^{3}$ \and
D.~J.~Price$^{3}$ \and
S.~Inoue$^{4}$
}
\institute{Institute for Solar Physics, Department of Astronomy, 
 Stockholm University, AlbaNova University Centre,
 106 91 Stockholm, Sweden
 \and 
 Astrophysics Research Centre, School of Mathematics and Physics, Queen’s
 University Belfast, BT7 1NN, Northern Ireland, UK
 \and 
 Department of Physics, University of Helsinki, P.O. Box 64, 00014, Helsinki, Finland
 \and 
 Institute for Space-Earth Environmental Research (ISEE), Nagoya University
 Furo-cho, Chikusa-ku, Nagoya, 464-8601, Japan
}

\date{}

\abstract
  {Obtaining the magnetic field vector accurately in the solar atmosphere is
  essential for studying changes in field topology during flares and to reliably
  model space weather.}
  {We tackle this problem by applying various inversion methods to a confined
  X2.2 flare that occurred in NOAA AR 12673 on September 6, 2017, and comparing
  the photospheric and chromospheric magnetic field vector with those from two
  numerical models of this event.
  }
   {We obtain the photospheric magnetic field from Milne-Eddington and
   (non-)local thermal equilibrium (non-LTE) inversions of \Hinode\ SOT/SP
   \FeI\,\,6301.5\,\AA\ and 6302.5\,\AA.
   The chromospheric field is obtained from a spatially-regularised weak field
   approximation (WFA) and non-LTE inversions of \CaIR\ observed with CRISP at
   the \SST.
   We investigate the field strengths and photosphere-to-chromosphere shear in
   field vector.
  }
  {The LTE- and non-LTE-inferred photospheric magnetic field components are
  strongly correlated across several optical depths in the atmosphere, with a
  tendency for stronger field and higher temperatures in the non-LTE inversions. 
  For the chromospheric field, the non-LTE inversions correlate well with the
  spatially-regularised WFA, especially in line-of-sight field strength and
  field vector orientation. 
  The photosphere exhibits coherent strong-field patches of over 4.5\,kG,
  co-located with similar concentrations exceeding 3\,kG in the chromosphere. 
  The obtained field strengths are up to 2--3 times higher than in the numerical
  models and the photosphere-to-chromosphere shear close to the
  polarity inversion line is more concentrated and structured.} 
  {In the photosphere, the assumption of LTE for \FeI\ line formation does not
  yield significantly different magnetic field results compared to non-LTE,
  while Milne-Eddington inversions fail to reproduce the magnetic field vector
  orientation where \FeI\ is in emission.
  In the chromosphere, the non-LTE-inferred field is excellently approximated by
  the spatially-regularised WFA.
  Our inversions confirm the locations of flux rope footpoints that are
  predicted by numerical models.
  However, pre-processing and lower spatial resolution lead to weaker and
  smoother field in the models than what the data indicate. 
  This emphasises the need for higher spatial resolution in the models to better
  constrain pre-eruptive flux ropes.
  }
\maketitle

\section{Introduction}\label{sec:introduction}
Flares and coronal mass ejections (CMEs) are the source of the most violent
heliospheric disruptions and understanding what triggers them is an essential
piece in the space weather puzzle.
To reliably predict either is, however, not straightforward. 
Current forecasting efforts 
build on the round-the-clock coverage of the Earth-facing side of the Sun by the
\SDO\ (SDO;
\citeads{2012SoPh..275....3P}) 
that provides a view from the photosphere to the corona in the (extreme)
ultraviolet, as well as photospheric magnetic field information from the
\FeI\,\,6173\,\AA\ line.
Such data allow for parametrisation of the photospheric field over large
regions, using properties of \eg\ the polarity inversion line 
\citepads{2007ApJ...655L.117S}, 
connectivity across it
\citepads{2007ApJ...661L.109G} 
and measures of the magnetic shear or helicity (\eg\
\citeads{2012ApJ...760...31K}, 
\citeads{2017A&A...601A.125P}, 
\citeads{2018ApJ...863...41Z}) 
that are known to be indicators of an active region's eruptive potential. 
These have been exploited by themselves or combined with SDO's upper-atmosphere
diagnostics to arrive at a prediction through either statistical models or
machine learning techniques that have become increasingly popular over the past
few years (\eg\ 
\citeads{2003ApJ...595.1296L}, 
\citeyearads{2007ApJ...656.1173L}, 
\citeads{2015ApJ...798..135B}, 
\citeads{2016SoPh..291.1711M}, 
\citeads{2018SoPh..293...28F}, 
\citeads{2018SoPh..293...48J}, 
\citeads{2018ApJ...858..113N}, 
\citeads{2020ApJ...891...17P}). 

However, this means that usually only the lower magnetic (and/or the derived
electric) field boundary conditions are taken into account,
while the chromospheric magnetic field vector
could significantly aid
data-driven modelling (\eg\
\citeads{2009ApJ...696.1780D}, 
\citeads{2020ApJ...890..103T}) 
of, for instance, the magnetic field structure of CMEs
\citepads{2019SpWea..17..498K}, 
which in turn is important for space weather forecasts.
An important obstacle is, however, that the necessary spectropolarimetric
observations are neither widely nor commonly acquired, and when they are the
field-of-view is typically limited.
Consequently, flare studies that include chromospheric polarimetry---and do so at
high spatial resolution---are sparse (\eg\
\citeads{2014A&A...561A..98S}, 
\citeads{2015ApJ...799L..25K}, 
\citeads{2015ApJ...814..100J}, 
\citeads{2017ApJ...834...26K}, 
\citeads{2018ApJ...860...10K}, 
\citeads{2019A&A...621A..35L}, 
\citeads{2019ApJ...874..126K}). 

Towards the end of the last Solar Cycle 24 NOAA active region (AR) 12673 evolved
from a lonely symmetric sunspot as it appeared when rotating into view to an
active and complex sunspot group by the time it crossed the central meridian.
The flaring activity of this active region has been studied extensively over the
past two years, given that over the span of a week it produced four X-class
flares---including the two largest flares of that cycle (X9.3 on September 6
11:53\,UT and X8.2 on September 10 15:35\,UT)---over two dozen M-class
flares and many more smaller ones.
Moreover, the X9.3 flare was preceded by a confined X2.2 flare a mere 3\,h
earlier.
The active region developed as flux that emerged over the course of three days
next to an $\alpha$-class sunspot coalesced and led to a complex $\delta$-spot
configuration, where strong shearing flows along the polarity inversion line
(PIL) between the positive-polarity spot and parasitic negative polarity 
were the likely main agent in setting up the active region for flaring (%
\citeads{2017ApJ...849L..21Y}, 
\citeads{2018ApJ...852L..10R}, 
\citeads{2018ApJ...869...90W}, 
\citeads{2018A&A...612A.101V}). 

With near-continuous coverage by the \HMI\ (HMI;
\citeads{2012SoPh..275..207S}, 
\citeads{2012SoPh..275..229S}) 
aboard SDO 
and availability of its derived data products such as Spaceweather HMI Active
Region Patches (SHARPs;
\citeads{2014SoPh..289.3549B}), 
this is also an attractive target to study the magnetic field configuration and
evolution during emergence and flaring. 
For instance, 
\citetads{2018A&A...619A.100H} 
focussed on the two largest flares that the active region brought forth (\ie\
the X9.3 and X8.2 flares) and used a time sequence of non-linear force free
field (NLFFF) extrapolations to investigate the field evolution during the
flares.
For both flares they found a multi-flux-rope configuration over the PIL 
wherein the flux ropes were destabilised by the shearing and rotating motions of
the $\delta$-sunspot, setting off the upper flux rope and leading to
destabilisation of adjacent flux ropes that ended up erupting shortly after.
Following a similar approach, 
\citetads{2018ApJ...867L...5L} 
used a time series of NLFFF and potential field models to study the confined
X2.2 flare that preceded the X9.3 one and they also identified a multiple-branch
or double-decker magnetic flux rope configuration.
During the confined flare the magnetic helicity was found to increase by over
250\% from pre-flare values and again significantly reduced during the
subsequent X9.3 flare, implying a scenario in which the confined flare set the
stage for the eruptive one.
\citetads{2019ApJ...870...97Z}, 
analysing a series of NLFFF extrapolations obtained using a magnetohydrodynamic
(MHD) relaxation method, suggested a two-step reconnection process in which
reconnection first occured in a null point outside the magnetic flux rope system
and at around 4000\,km height, while the associated disturbance then triggered
the second, tether-cutting reconnection.
However, as the overlying field was sufficiently strong, the flare remained
confined.
\citetads{2019SoPh..294....4R} 
performed NLFFF extrapolations as well and identified null points for the X2.2 and
X9.3 flares at 5000 and 3000\,km above the photospheric boundary, respectively.

More elaborate modelling was performed by 
\citetads{2018ApJ...867...83I}, 
\citetads{2018ApJ...869...13J}, 
and
\citetads{2019A&A...628A.114P}. 
The first performed a MHD simulation for which the initial magnetic
configuration was set by a NLFFF extrapolation from a SHARP about 20\,min prior
to the X2.2 flare. 
Their results suggest that the X2.2 flare may have been associated with the rise
of a small flux rope, triggered by reconnection underneath it and that the X9.3
flare was likely to be the eruption of a large-scale magnetic flux rope that had
been formed through reconnection between several smaller ones in the hours
leading up to the large flare.
Similarly, 
\citetads{2018ApJ...869...13J}, 
analysed a NLFFF extrapolation-initialised MHD simulation of the X9.3 flare
and identified tether-cutting reconnection as the likely trigger of the flux rope
eruption.
On the other hand, \citetads{2019A&A...628A.114P} 
performed time-dependent magnetofrictional modelling to investigate the
emergence and evolution that led up the X9.3 flare and associated eruption and
coronal mass ejection. 
As such, their simulation covers also the preceding confined X2.2 flare.
Feeding the model with a time series of electric field inversions based on the
SDO/HMI vector magnetic field, with the magnetic field initialised from a
potential field extrapolation, the authors report an increase in helicity
during the X2.2 flare consistent with the findings of
\citetads{2018ApJ...867L...5L} 
and while the magnetic flux rope did not erupt out of the simulation's numerical
domain during the X9.3 flare, in contrast to the results by
\citetads{2018ApJ...867...83I}, 
both studies produce a similar large-scale field configuration and evolution, as
well as structure of the erupting flux rope.

However, a limitation of all above field extrapolation and modelling efforts
remains the lack of chromospheric magnetic field input.
Including chromospheric field information has been shown to aid the NLFFF
extrapolation in recovering the chromospheric and coronal magnetic field
structure
\citepads{2019ApJ...870..101F}. 
Similarly,
\citetads{2020ApJ...890..103T} 
compared several data-driven model results based on a flux emergence simulation
and found that the largest errors were due to \rev{strong Lorentz forces at the
photospheric boundary inducing spurious flows that altered the magnetic field
(via the induction equation)},
whereas inclusion of the field at a higher\rev{, and much more force-free,}
layer resulted in better agreement with the input simulation.
While the (non-magnetic) \Halpha\ response to the X9.3 flare has been studied in
detail in
\citetads{2019ApJ...881...82Q}, 
the chromospheric magnetic field of neither of the September 6 flares has
previously been investigated from observations.

We take on part of that challenge in the present study and 
focus on the confined X2.2 flare in NOAA AR 12673 
for which high-resolution observations of both photospheric
and chromospheric spectropolarimetry are available.
Section~\ref{sec:observations} introduces the observations and post-processing
to prepare the data for inversions.
Section~\ref{sec:methods} describes different approaches we employed in
inferring the photospheric and chromospheric magnetic field vectors, with
Section~\ref{sec:results} presenting our results. 
We also compare those against two numerical models in
Section~\ref{sec:results_inv_model}.
We discuss our findings in Section~\ref{sec:discussion} and present our
conclusions in Section~\ref{sec:conclusions}.

\section{Observations and reduction}\label{sec:observations}
\begin{figure*}[bht]
  \centerline{\includegraphics[width=\textwidth]{\figspath/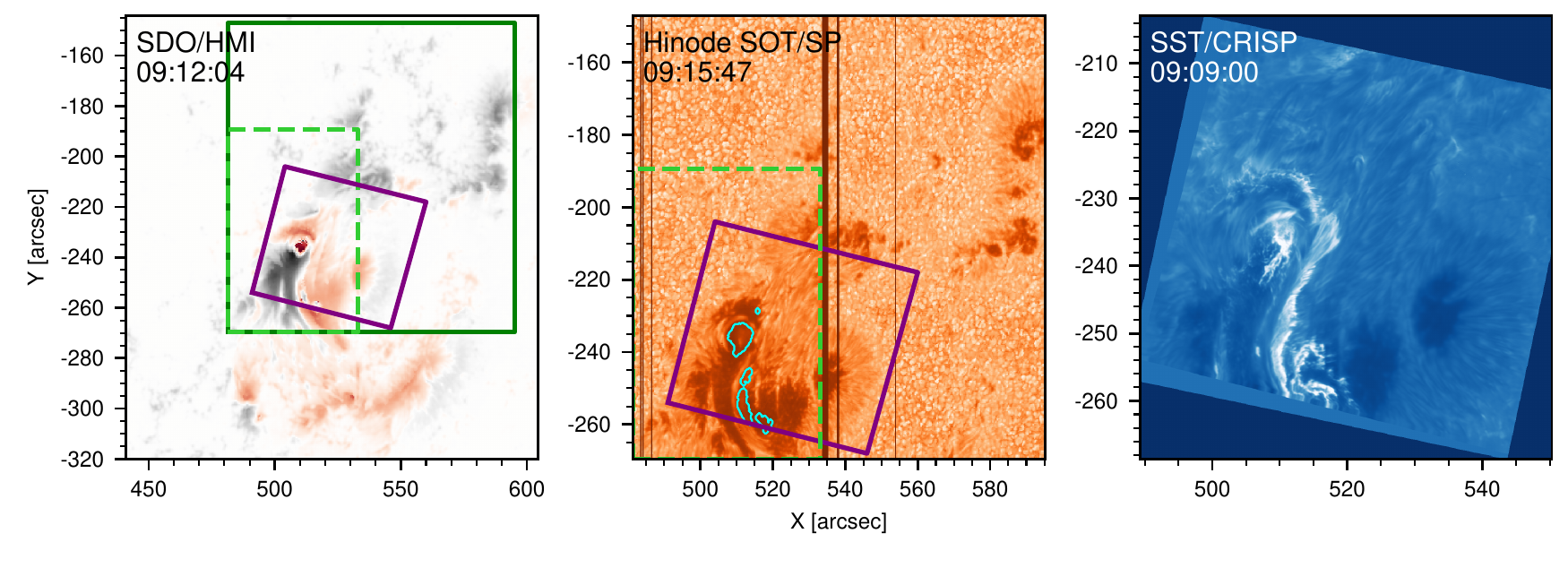}}
  \vspace{-4.5ex}
  \caption[]{\label{fig:fovs} %
    Overview of the analysed data, showing the SDO/HMI-derived SHARP
    line-of-sight magnetic field from \FeI\,\,6173\,\AA\ ({\it left\/}),
    \Hinode/SP continuum near \FeI\,\,6302\,\AA\ ({\it middle\/}) and SST/CRISP
    \CaIR\ red wing at +0.5\,\AA\ ({\it right\/}).
    The \Hinode\ and SST fields-of-view are indicated in the first two panels with
    green and purple boxes, respectively, while the dashed light green box in the
    first two panels outlines the cut-out of Fig.~\ref{fig:fe_fields}. 
    The cyan contours in the middle panel indicate locations where the \FeI\
    lines are in emission.
    Times in UT are indicated in the top left of each panel (for \Hinode\ SOT/SP the
    time corresponds to the middle of the slit raster-scan). 
    The vertical stripes in the middle panel are due to missing data.
  }
\end{figure*}

We analyse a confined X2.2 flare in NOAA AR 12673 on September 6, 2017, that
lasted from 08:57--09:17\,UT, peaking at 09:10\,UT.
Part of its rise and decay phase was observed by the Solar Optical Telescope
(SOT;
\citeads{2008SoPh..249..167T}) 
Stokes Spectro-Polarimeter (SP) aboard \Hinode\ 
\citepads{2007SoPh..243....3K}, 
as well as 
the CRisp Imaging SpectroPolarimeter (CRISP; 
\citeads{2008ApJ...689L..69S}) 
at the Swedish 1-m Solar Telescope (SST;
\citeads{2003SPIE.4853..341S}). 
Figure~\ref{fig:fovs} presents an overview of these observations, along with
context line-of-sight magnetic field from a SDO/HMI-derived SHARP two minutes
after the flare peak.

Imaging spectropolarimetry in the \CaIR\ line was obtained at the SST between
09:04:30--09:54:24\,UT, sampling 11 wavelength positions out to $\pm$0.7\,\AA\
from line centre (at 0.1\,\AA\ spacing between $\pm$0.3\,\AA\ and at 0.2\,\AA\
in the wings) at an overall cadence of 15\,s per scan.
The pixel scale is 0\farcs{058}\,pix$^{-1}$.
The data were reduced with the CRISPRED 
\citepads{2015A&A...573A..40D} 
pipeline, including image restoration using Multi-Object Multi-Frame Blind
Deconvolution (MOMFBD;
\citeads{2005SoPh..228..191V}), 
removal of remaining small-scale seeing-induced deformations
\citepads{2012A&A...548A.114H} 
and destretching to correct for rubber-sheet seeing effects
\citepads{1994ApJ...430..413S}. 

Additional treatment of the data was necessary to increase the signal-to-noise
in \CaIR\ Stokes Q and U, and we applied the denoising neural network of
\citetads{2019A&A...629A..99D}, 
followed by Fourier-filtering to suppress the strongest remaining high-frequency
fringe patterns (similar to the method described in \citetads{2020arXiv200614486P}).
As the seeing conditions were variable at La Palma, we selected a single line
scan snapshot at 09:09:00\,UT for further study, based on best contrast
throughout the line scan and fortuitously close in time to the flare peak. 

\Hinode\ SOT/SP performed a fast map of \fov{164}{164} full spectropolarimetry
in \FeI\,\,6301.5\,\AA\ and 6302.5\,\AA\ at 3.2\,s integration time per slit
position between 09:03:40--09:27:53\,UT, resulting in a mid raster-scan time of
09:15:47\,UT.
The raster pixel size is $0\farcs{32} \times 0\farcs{30}$.
For the sub-field cutout marked by the dashed box in Fig.~\ref{fig:fovs}, the
mid-scan time is 09:09:14\,UT, close to the selected best-contrast frame of the
SST data set and the flare peak.
We did not attempt simultaneous inversion of the \Hinode\ and SST data,
as SOT/SP required \rev{nearly} 11\,min to scan the sub-field, leading to
increasingly inconsistent \FeI\ and \CaII\ profiles towards the vertical edges
of the overlapping field-of-view.
\rev{However, the SOT/SP sampling of the polarity inversion line
vicinity---which is what we are primarily interested in---falls within 3\,min of
the \CaII\ line scan snapshot.}

\subsection{Data alignment}\label{sec:alignment}
We aligned the SST data to \Hinode\ through cross-correlation, using the
\Hinode\ continuum image (Fig.~\ref{fig:fovs}, middle panel) as anchor to which
the \CaIR\ wide-band image was aligned.
This includes downsampling the SST
data to \Hinode\ resolution (about a factor 5 in both $x$ and $y$) as part of
the alignment process.
As we are primarily interested in comparing the photospheric and chromospheric
field in the same pixels, this is a reasonable compromise to make and it has the
added benefit of saving computational time, as well as improving the
signal-to-noise ratio of the polarimetric data.
CRISPEX (%
\citeads{2012ApJ...750...22V}, 
\citeads{2018arXiv180403030L}) 
was used to verify inter-instrument alignment and for data browsing.

\section{Methods for magnetic field vector inference}\label{sec:methods}
Several methods exist to infer or derive the magnetic field vector from
observations, varying in degree of complexity and computational expense. 
In this section we discuss the three methods used in this study,
with on one end Milne-Eddington inversions and a spatially-regularised
weak-field approximation that provide a field estimate under simplifying
assumptions and on the other end (non-)LTE inversions that yield a
depth-stratified model atmosphere with temperature, velocities and magnetic
field, albeit at higher computational cost.

\subsection{Milne-Eddington inversions} 
As approximation of the photospheric field we use the pixel-by-pixel (\ie\ each
pixel is treated independently) Milne-Eddington (ME) inversion results obtained
with the \rev{Milne-Eddington gRid Linear Inversion Network (MERLIN)} code.
These inversions are performed assuming a source function that is linear
with optical depth, but where quantities like the magnetic field vector and
line-of-sight velocity are otherwise constant throughout the atmosphere.
We also note that in the routine application of the MERLIN code a saturation limit of
5\,kG is imposed on the line-of-sight and horizontal field strengths.
The SOT/SP level-2 data product readily delivers these results and contains
(among other quantities) the field strength value, its inclination and azimuth,
where the latter still has an unresolved 180\deg-ambiguity.
We discuss our disambiguation approach for this and the other inferred azimuths
in Section~\ref{sec:disambig}.

\subsection{Spatially-regularised weak-field approximation} 
An estimate of the chromospheric magnetic field can be obtained using the
weak-field approximation (WFA, %
\citeauthor{1992soti.book...71L} 
\citeyearads{1992soti.book...71L}, 
\citeyearads{2004ASSL..307.....L}), 
which assumes that the Zeeman splitting is smaller than the Doppler broadening
of the line in question.
This method does not bear the cost of solving the full non-LTE radiative
transfer problem and lends itself therefore well for fast estimation of the
magnetic field over a larger field-of-view, but suffers from several
simplifications and a limited range of validity that need be borne in mind
\citepads{2018ApJ...866...89C}. 
The WFA has found its use for targets like sunspots
\citepads{2013A&A...556A.115D}, 
plage
\citepads{2007ApJ...663.1386P}, 
as well as in flares (%
\citeads{2012SoPh..280...69H}, 
\citeads{2017ApJ...834...26K}). 

Here we use a spatially-regularised WFA 
\citepads{2020arXiv200614487M}---
an extension of the commonly used pixel-by-pixel one---to infer the approximate
chromospheric magnetic field configuration over the full SST field-of-view.
The spatially-regularised approach departs from the idea that,
when the observations are properly sampled near the diffraction limit of the
telescope, the derived magnetic field should be spatially smooth in its
variation.
The implementation we use employs Tikhonov $\ell$-2 regularisation and to
impose the smoothness the values of the four nearest-neighbours ($\pm$1 pixel in
both the $x$- and $y$-direction) are taken into account when minimising \chisq.
The power of this method is that with well-chosen parameters the effects of
noise can be drastically mitigated.
For further details we refer the reader to
\citetads{2020arXiv200614487M}.  

Finally, we note that the weak-field approximation results (including azimuth
disambiguation) were obtained before downsampling to \Hinode\ resolution was
performed as part of the data alignment process.

\subsection{Non-LTE inversions}
We use the \stic\ 
(STiC;
\citeads{2016ApJ...830L..30D}, 
\citeads{2019A&A...623A..74D}) 
to infer the atmospheric stratification of temperature,
velocities and magnetic field from the \Hinode\ \FeI\ and SST \CaII\ data.
STiC is an MPI-parallel non-LTE inversion code built around a modified version
of RH 
\citepads{2001ApJ...557..389U} 
to solve the atom population densities assuming statistical equilibrium and
plane-parallel geometry, using an equation of state extracted from the SME code 
\citepads{2017A&A...597A..16P}. 
We assumed complete frequency redistribution (CRD) in our inversions.
The radiative transport equation is solved using cubic Bezier solvers 
\citepads{2013ApJ...764...33D}. 
As the \Hinode\ SOT/SP scanning time was too long to obtain consistent \FeI\ and
\CaII\ profiles for the overlapping part of the field-of-view, we decided to
perform the inversions for both lines separately and could therefore not 
take advantage of the multi-resolution inversion technique recently proposed by 
\citetads{2019A&A...631A.153D}.  

The \CaII\ inversions were performed in non-LTE using a 6-level calcium model
atom and assuming CRD.
For \FeI\ we performed both LTE and non-LTE inversions, spurred in part by the 
recent study by 
\citetads{2020A&A...633A.157S}.  
They report on LTE inversions of \FeI\ 6301\,\AA\ and 6302\,\AA\ that were
synthesised assuming either LTE or non-LTE and that indicate that LTE inversions
of non-LTE \FeI\ may result in discrepancies with the input model in terms of
temperature (of order 10\%) and line-of-sight velocities and magnetic field
(both up to 50\%), while the field inclination could exhibit errors of up to
45\deg.
For both the LTE and non-LTE \FeI\ inversions we used the same extended 23-level
\FeI\ atom that has been used in several studies on the non-LTE radiative
transfer effects in the \FeI\ lines over the past decade (%
\citeads{2013A&A...558A..20H},  
\citeyearads{2015A&A...582A.101H},  
\citeads{2020A&A...633A.157S}).  

As initial input atmosphere we used a FAL-C model interpolated to a $\dltau=0.2$
grid between $\ltau=-8.5$ and 0.1, but truncated at $\ltau=-5$ (\ie\ excluding
the upper atmosphere) for \FeI\ given the lack of sensitivity of the line to
conditions much above the temperature minimum.
Both the \FeI\ and \CaII\ inversions were performed in two cycles with spatial
and depth smoothing of the model atmosphere between the first and second cycle,
as well as an increased number of nodes in temperature, velocity and magnetic
field component in the second cycle (see Table~\ref{tab:cycles}).
In addition, as the flaring emission profiles in both \FeI\ and \CaII\ proved a
challenge for the first cycle inversions, we replaced the atmospheres of the
worst fitted pixels (as expressed by their \chisq-value) with the average
atmosphere of nearby well-fitted pixels with same-sign line-of-sight field prior
to smoothing for the second cycle input.
The results of these inversions are presented in Section~\ref{sec:results}.

\begin{table}[h]
\caption{Number of nodes used in each inversion cycle.}
\begin{center}
  
\begin{tabular}{l|ll|cccccc}%
  \hline\hline
  \multicolumn{3}{r|}{Parameter} & $T$ & \vlos & \vturb & \Blos & \Bhor & $\varphi$ \\
  \hline
  \multirow{4}{*}{\rotatebox[origin=c]{90}{Inversion}} & 
  \multirow{2}{*}{\FeI}   & cycle 1 & 4 & 1 & 0 & 1 & 1 & 1 \\
  {} & {}                 & cycle 2 & 5 & 2 & 1 & 2 & 2 & 1 \\ \cline{2-9}
  {} & \multirow{2}{*}{\CaII}  & cycle 1 & 4 & 1 & 0 & 1 & 1 & 1 \\
  {} & {}                 & cycle 2 & 7 & 2 & 1 & 2 & 2 & 1 \\
  \hline
\end{tabular}
\end{center}
\label{tab:cycles}
\end{table}

\subsection{Azimuth disambigution}\label{sec:disambig}
The photospheric and chromospheric magnetic field azimuth $\varphi$ recovered from both
the Milne-Eddington inversions, the weak-field approximation and STiC inversions
contain a 180\deg-ambiguity that needs to be resolved for proper interpretation
of the horizontal magnetic field component.
Several schemes exist for this disambiguation
\citepads{2006SoPh..237..267M};  
here we use the implementation by 
\citetads{2014ascl.soft04007L}  
of the minimum energy method (MEM) proposed by
\citetads{1994SoPh..155..235M},  
that simultaneously minimises the divergence of the field and the current
density.
This method works well for photospheric lines, where the linear polarisation
signal is sufficiently strong, but the method typically struggles for \CaIR\
where the Stokes Q and U profiles are noisier.
Above the sunspot and during the flare, however, the linear polarisation signal
is sufficiently strong that the azimuth can be reliably recovered in the
chromosphere.
In order to check what areas of the FOV are uncertain for the ambiguity
resolution, we ran the MEM code with twenty different random number seeds.
Regions where the azimuth result changed by more than 45\deg\ between runs were
considered to be unstable for disambiguation and for those pixels we set the
azimuth by majority vote of the results from the twenty realisations.

\section{Observationally inferred magnetic field vector}
\label{sec:results}

\begin{figure*}[bht]
  \centerline{\includegraphics[width=\textwidth]{\figspath/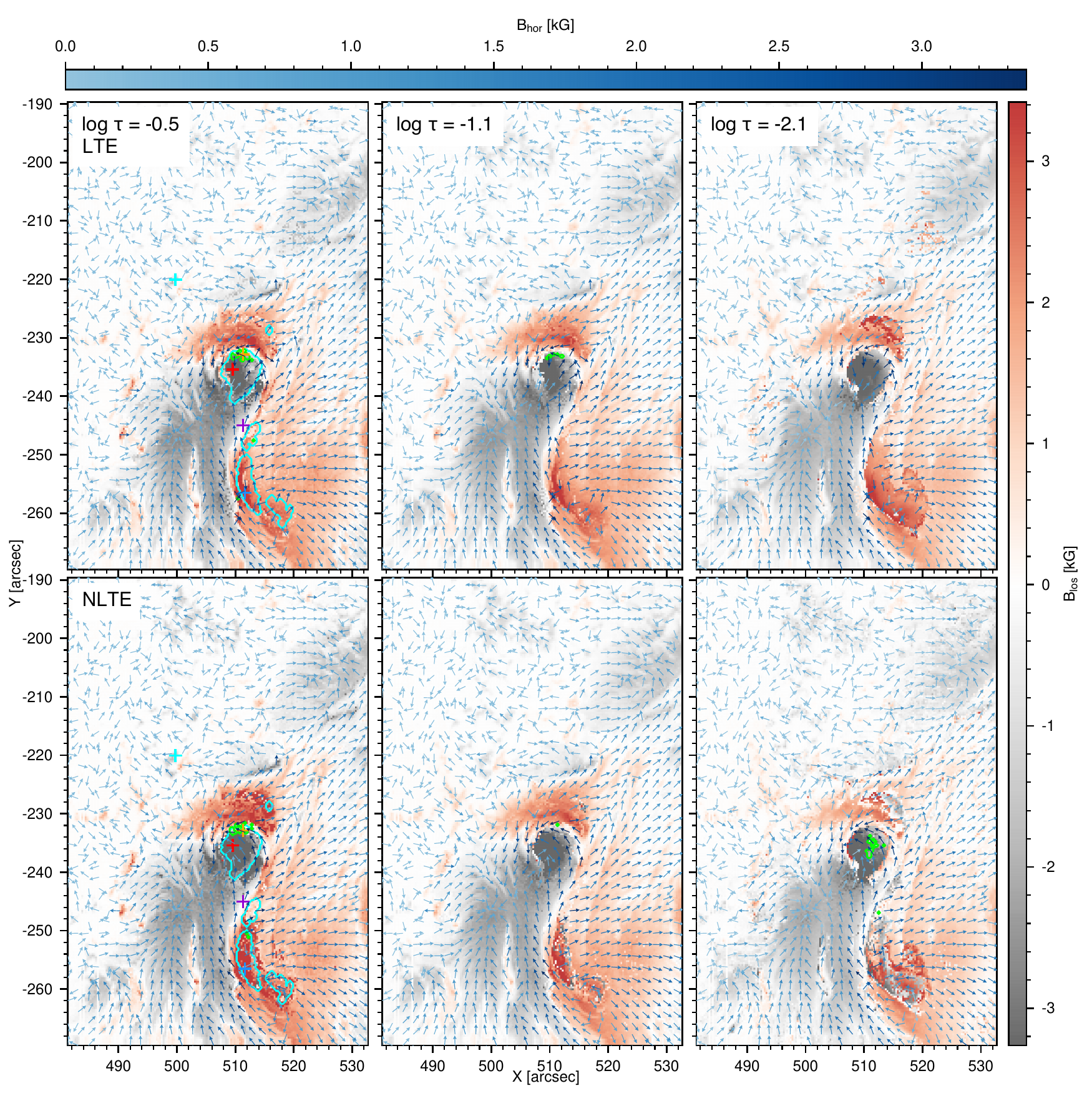}}
  \vspace{-2.5ex}
  \caption[]{\label{fig:fe_fields} %
    Line-of-sight and transverse magnetic field from STiC (non-)LTE inversions of the \FeI\ \Hinode\ data over the
    sub-field indicated by the dashed green box in Fig.~\ref{fig:fovs}.
    From left to right the columns show the photospheric field at
    different \ltau\ depths as specified in the top left corner of each panel,
    both for LTE ({\it top row\/}) and non-LTE ({\it bottom row\/}) inversions.
    The maps are of the line-of-sight field (scaled according to the right-hand
    colour bar), while the blue arrows indicate the azimuth direction, where
    their hue reflects the horizontal field strength (according to the top
    colour bar).
    All panels are colour-scaled between the same values with both line-of-sight
    and horizontal field strengths clipped to the the range wherein 98\% of the
    pixels fall.
    Bright green contours indicate where the horizontal field is in excess of
    5\,kG, while the cyan contours in the first column highlight where the \FeI\
    lines are in emission.
    The coloured plus signs mark the locations for which (non-)LTE profile fits
    are shown in Fig.~\ref{fig:profs}.
    }
\end{figure*}

\begin{figure*}[bht]
  \centerline{\includegraphics[width=\textwidth]{\figspath/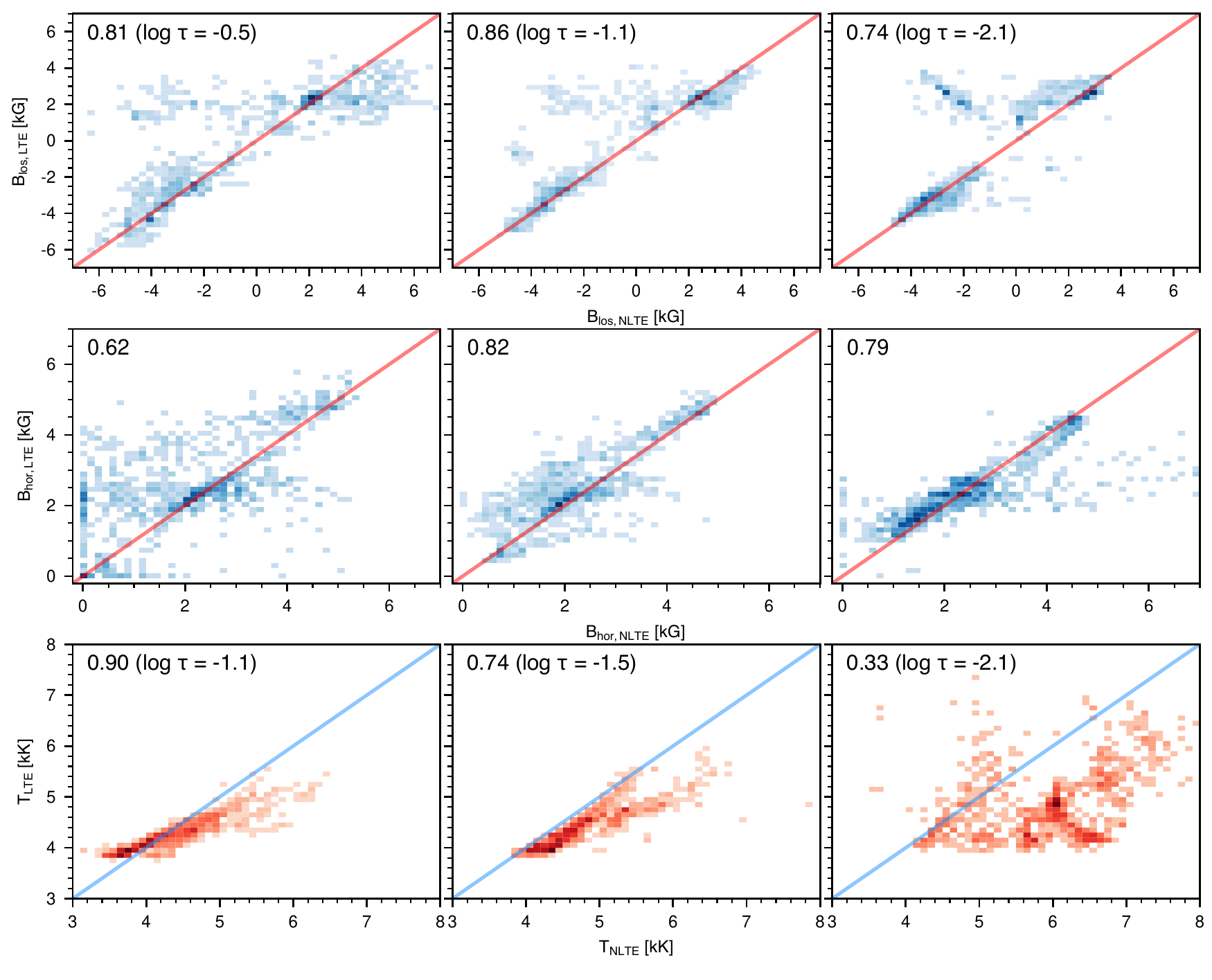}}
  \vspace{-3.5ex}
  \caption[]{\label{fig:scatter} %
    Two-dimensional histograms of non-LTE versus LTE line-of-sight magnetic
    field strength ({\it top row\/}), horizontal field strength ({\it middle
    row\/}) and temperature ({\it bottom row\/}) for the flaring pixels (\ie\
    within cyan contours in Fig.~\ref{fig:fe_fields}) at three \ltau\ depths.
    The Pearson correlation number is shown in the top left of each panel and
    the \ltau\ depth within parantheses in the top row panels (for magnetic
    field) and bottom row panels (for temperature), while the straight lines
    indicate what would be linear relationship.
    }
\end{figure*}

\begin{figure*}[bht]
  \centerline{\includegraphics[width=\textwidth]{\figspath/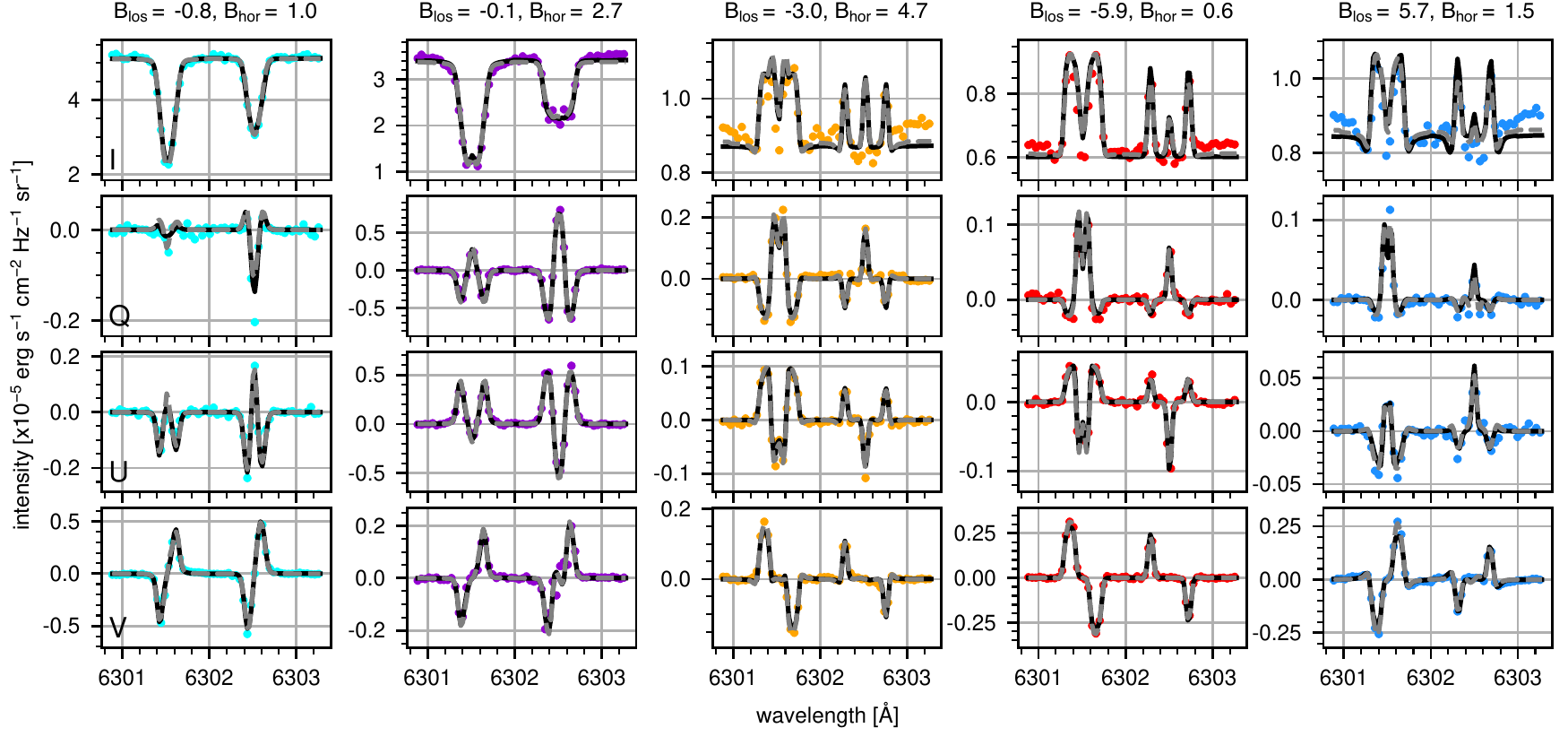}}
  \vspace{-2ex}
  \caption[]{\label{fig:profs} %
    \FeI\,6301.5\,\AA\ and 6302.5\,\AA\ profiles from observations and (non-)LTE
    inversions.
    Each column shows ({\it from top to bottom\/}) Stokes $I$, $Q$, $U$ and $V$ 
    profiles as observed ({\it coloured dots\/}) and as fitted in LTE
    ({\it grey dashed line\/}) and in non-LTE ({\it solid black line\/}).
    The colour coding corresponds to the identically coloured plus markers in the
    left-hand panels of Fig.~\ref{fig:fe_fields}.
    The numbers above each column indicate the non-LTE-inferred \Blos\ and
    \Bhor\ at $\ltau=-0.5$.
    }
\end{figure*}

\begin{figure*}[bht]
  \centerline{\includegraphics[width=\textwidth]{\figspath/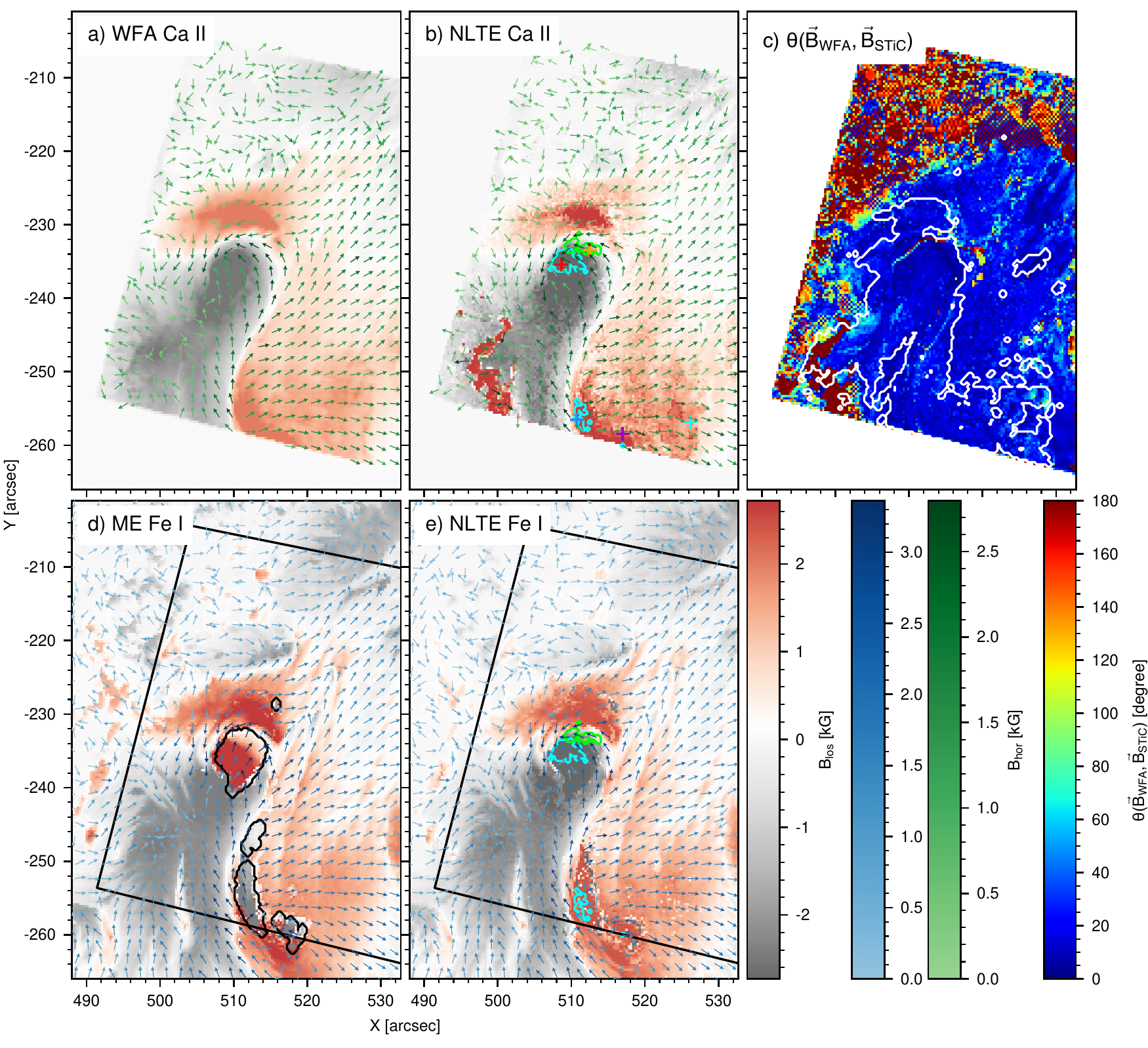}}
  \vspace{-2.5ex}
  \caption[]{\label{fig:mewfa_stic} %
    \rev{Chromospheric and photospheric} magnetic field from the considered
    inversion methods.
    \rev{%
      {\it Panel (a)\/}: Chromospheric line-of-sight field maps with green azimuth arrows
      coloured according to its horizontal field strength from the
      spatially-regularised WFA. 
      {\it Panel (b)\/}: Same as (a) but from STiC non-LTE inversions.
      The contours mark where the field exceeds 4.5\,kG in the photosphere
      and 3\,kG in the chromosphere for \Blos\ ({\it cyan\/}) and the same
      thresholds for \Bhor\ ({\it light green\/}).
      The coloured plus markers indicate the locations for which
      Fig.~\ref{fig:profs_ca} shows \CaIR\ profile fits.
      {\it Panel (c)\/}: Angle difference
      $\theta(\vec{B}_{\rm{WFA}},\vec{B}_{\rm{STiC}})$ between the WFA and
      non-LTE three-dimensional field vectors, clipped to 100\deg.
      The dashed white contours highlight the regions where the WFA field
      strength exceeds 2\,kG (selected pixels for Fig.~\ref{fig:ca_scatter}).
      {\it Panel (d)\/}: Photospheric line-of-sight field maps with blue azimuth
      arrows as derived in the Milne-Eddington inversion.
      The black box indicates the SST field-of-view, while the contours
      highlight where the \FeI\ lines are in emission.
      {\it Panel (e)\/}: Same as (d) but from the non-LTE inversions and contours
      as in panel (b).
    The colour bars in the lower right are for the line-of-sight field
    (grey-white-red), horizontal field in the photosphere (blue) and
    chromosphere (green), and the angle $\theta$ (rainbow).
    The colour bar range for the magnetic field strengths is representative
    of 98\% of the pixels (as in Fig.~\ref{fig:fe_fields}).}
  }
\end{figure*}

\begin{figure*}[bht]
  \centerline{\includegraphics[width=\textwidth]{\figspath/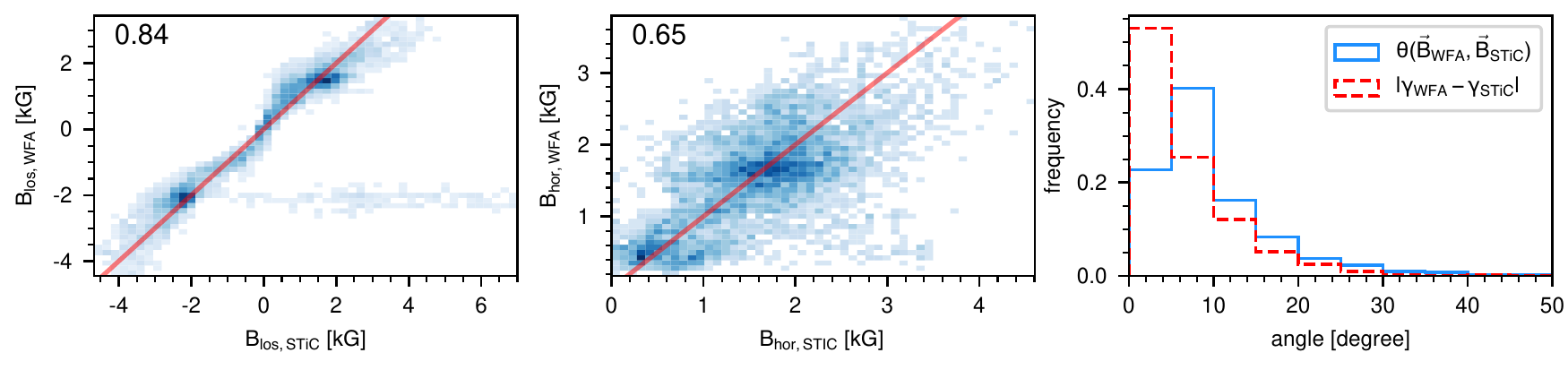}}
  \vspace{-2.5ex}
  \caption[]{\label{fig:ca_scatter} %
    Two-dimensional histograms of non-LTE versus WFA line-of-sight magnetic field
    strength ({\it left\/}), horizontal field strength ({\it middle\/}) and
    histograms of angle differences ({\it right\/}) between the
    three-dimensional field vector as recovered from the WFA and non-LTE STiC
    inversions ({\it solid blue\/}) and between their field inclinations
    $\gamma_{\rm{WFA}}$ and $\gamma_{\rm{STiC}}$ ({\it dashed red\/}) for pixels where the total WFA magnetic field
    strength exceeds 2\,kG (white contours in Fig.~\ref{fig:mewfa_stic}\rev{d}). 
    Format for the left two panels as in Fig.~\ref{fig:scatter}.
    }
\end{figure*}

\begin{figure*}[bht]
  \centerline{\includegraphics[width=\textwidth]{\figspath/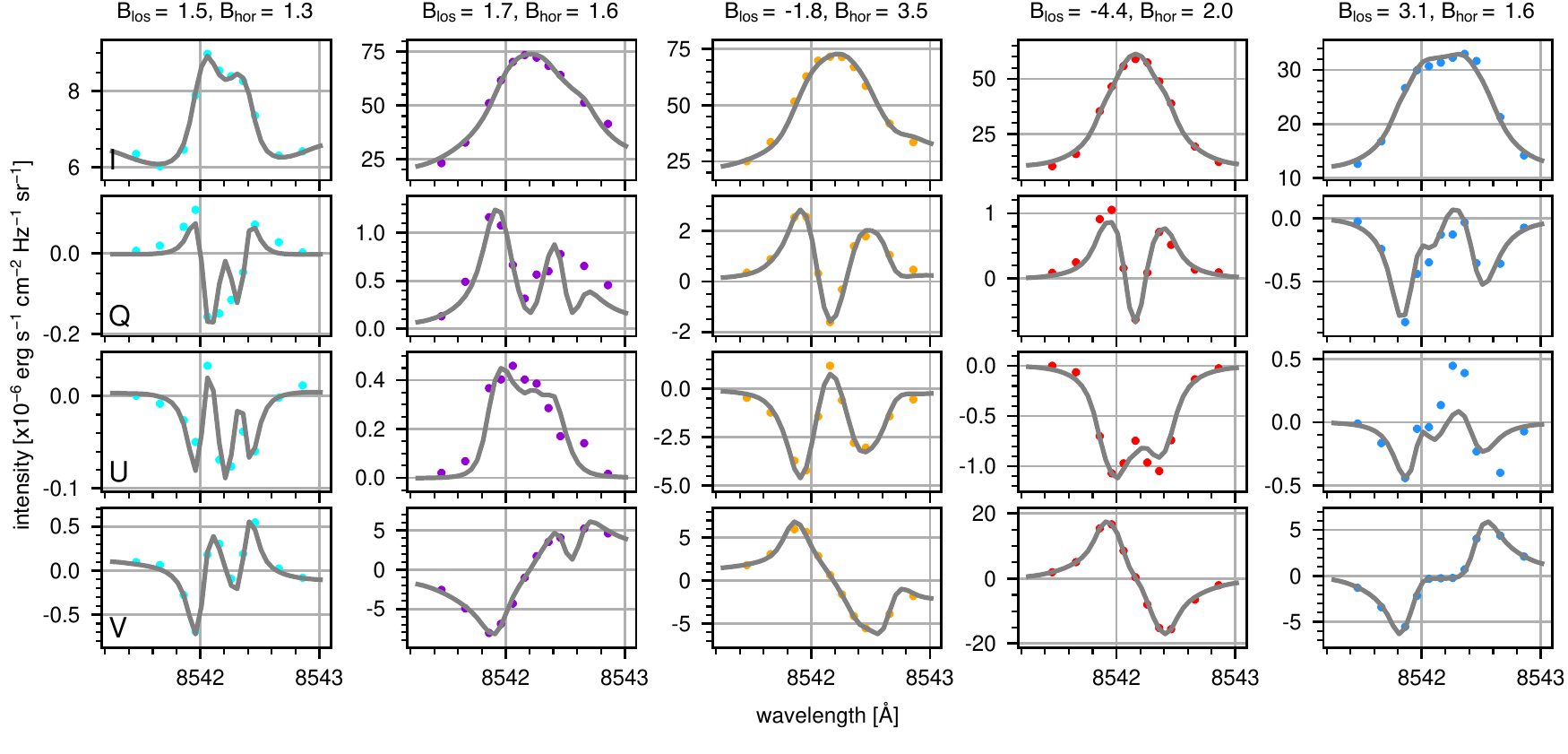}}
  \vspace{-2ex}
  \caption[]{\label{fig:profs_ca} %
    \CaIR\ profiles from observations and non-LTE inversions for the coloured
    plus markers in Fig.~\ref{fig:mewfa_stic}b.
    Format as for Fig.~\ref{fig:profs}, except that only non-LTE fitted profiles
    are shown ({\it solid grey\/}).
    The numbers above each column indicate the inferred \Blos\ and
    \Bhor\ averaged over $\ltau=[-2.9,-4.1]$.
    }
\end{figure*}

\subsection{LTE versus non-LTE inversions of \FeI}
\label{sec:results_inv}
Figures~\ref{fig:fe_fields}--\ref{fig:profs} compare the results from
the STiC LTE and non-LTE inversions of the \FeI\,\,6301.5\,\AA\ and 6302.5\,\AA\
spectra.
Figure~\ref{fig:fe_fields} shows the photospheric line-of-sight magnetic field
with arrows indicating the magnetic field azimuth where their hue reflects the
horizontal field strength (\ie\ darker blue being stronger).
Qualitatively, the LTE and non-LTE inversions of \FeI\ yield very
similar magnetic field distributions, especially at $\ltau=-0.5$ and $-$1.1.
Also the field azimuth shows largely the same pattern, as does to a certain
extent the strong horizontal field distribution (bright green contours for
$\Bhor > 5$\,kG).
Where the field is weak (both line-of-sight and horizontal, \eg\ in the top
left part of the field-of-view) the azimuth displays apparently random
orientations, contrasting with the more ordered pattern in the
strong-field sunspot umbra and penumbra, even though the distinct
`whirlpool'-like pattern in the `head' of the inverse-S shaped polarity inversion line
(\ie\ around \XYis{510}{-235}) is only recovered with the non-LTE inversions.
Regardless of LTE or non-LTE, the azimuth is essentially parallel to the
PIL in its vicinity, except in the positive-polarity
umbra around \XYis{515}{-245}.
The largest differences are primarily found at and close to the locations where
the \FeI\ lines go into emission, highlighted by the cyan contours in the first
column.
While both inversions yield negative polarity in the cyan-outlined `head' of
the inverse-S and
positive polarity in the western umbra, this only persists at all three shown
\ltau-depths for the LTE inversion (albeit no longer as smoothly at $\ltau=-2.1$),
while for the non-LTE inversion opposite polarity starts appearing around
\XYis{515}{-255} at $\ltau=-1.1$ and $-$2.1.

Figure~\ref{fig:scatter} quantifies this behaviour through two-dimensional
histograms of the line-of-sight and horizontal field, as well as the
temperatures for pixels where \FeI\ is in emission.
As the magnetic field scatter plots (top two rows) show, the correlation is
overall positive for \Blos\ and \Bhor\ and relatively tight for the former,
especially around $\ltau=-1.1$, even though there is a cloud of points at
negative \Blosnlte\ and positive \Bloslte\ at all three \ltau-depths.
Upon closer inspection, these turn out to be scattered pixels in the emission
patches around $Y=-250$\arcsec\ for which negative \Blos\ was inferred in
non-LTE.
These panels also evidence that, while in general neither \Blos\ nor \Bhor\
exceed 3\,kG by much (cf.~also the colour bars in
Fig.~\ref{fig:fe_fields}, clipped to the range wherein 98\% of the pixels fall),
there are pixels where values of 4--6\,kG are reached for either field
component.
Many pixels are also inverted with somewhat lower line-of-sight field strength
in LTE than in non-LTE (cf.~the scatter clouds above the diagonal line at
negative \Blosnlte\ and below at positive \Blosnlte\ for $\ltau=-0.5$ and
$-$1.1).
The opposite appears true at $\ltau=-2.1$ (top right panel).
The median fractional difference $(\Blosnlte-\Bloslte)/\Bloslte$ reaches up to
6.7\% at $\ltau=-0.5$, but decreases to $-$2.1\% at $\ltau=-2.1$, while for the
full field-of-view it peaks to 9.1\% at $\ltau=-1.1$.
On the other hand, for the horizontal field the LTE-inferred \Bhor\ is typically
stronger than in the non-LTE inversion, in particular at low and middle
\ltau-depths (median fractional difference of around $-$7\%), while at
$\ltau=-2.1$ pixels can be found above 4\,kG in non-LTE that barely reach that
value in the LTE inversion.

The discrepancies are even larger for temperature (bottom row of
Fig.~\ref{fig:scatter}), with strong correlation between LTE and
non-LTE-inferred temperatures at $\ltau=-0.5$, but an increasingly loose scatter
higher in the atmosphere at $\ltau=-1.5$ and especially $-$2.1.
At those heights fewer and fewer pixels lie on the diagonal and there is a clear
tendency for higher temperatures in non-LTE, inferring some
7--8\,kK versus 4--5\,kK in LTE.
The median non-LTE-to-LTE fractional difference increases from 1.5\% at
$\ltau=-0.5$ to nearly 25\% at $\ltau=-2.1$ for the flaring pixels.
For the full field-of-view that difference reaches 5.6\% at the same height.

Finally, Fig.~\ref{fig:profs} presents several examples of observed profiles and
LTE and non-LTE fits to those, one from a magnetic concentration outside the
sunspot and four from the (vicinity of the) flare.
Striking for some of the latter is the evident Zeeman splitting in
\FeI\,\,6302.5\,\AA\ Stokes $I$, suggesting strong magnetic fields.
As these panels show, we find only very minor differences between the LTE and
non-LTE fits, regardless of the profile shape.
In the ``quiet Sun'' sampling (first column) the 6301.5\,\AA\ Stokes $Q$
is better fitted in LTE, while the non-LTE fit to 6302.5\,\AA\ gets closer to
the observations.
The PIL-sampling (second column) is similarly fitted in either approach, but
it is in fact possible to fit the flaring emission profiles (last three columns)
even in LTE, with little difference compared to non-LTE.
Especially Stokes $V$ is well-fitted in these cases and both LTE and non-LTE
sometimes struggle in reaching the extrema (\eg\ 6302.5\,\AA\ Stokes $U$ in the
middle sampling (orange) or 6301.5\,\AA\ Stokes $Q$ in the last (blue) one).
Either way, the outer wings of the \FeI\ lines in Stokes $I$ generally prove
to be challenging to fit simultaneously with the emission peaks, where the LTE
fit is marginally closer (yet still not close) to the observations.

\subsection{Milne-Eddington and weak-field approximation versus non-LTE inversions}
\paragraph{Photospheric field}
\label{sec:results_me}
Figure~\ref{fig:mewfa_stic} compares the Milne-Eddington photospheric and
weak-field approximation chromospheric fields (left panel\rev{s (a) and (d)}) with their
equivalents from non-LTE inversions (middle panel\rev{s (b) and (e)}).
Let us first consider the photospheric field (\rev{bottom panels (d) and (e)}).
A limitation inherent to the Milne-Eddington inversions is that a linear source
function cannot simultaneously explain absorption and emission features.
Considering that where the \FeI\ cores are in emission the wings generally
exhibit absorption dips (cf.~Fig.~\ref{fig:profs}), the Milne-Eddington
inversion will likely fit the wings under the assumption that the polarity
reversal in Stokes $V$ is due to a change in the sign of the line-of-sight
magnetic field, rather than a change in slope of the source function.
This explains the evident discrepancy in line-of-sight magnetic field where the
\FeI\ lines are in emission (\rev{black} contours).
While the Milne-Eddington inversion returns an embedded opposite polarity in
those areas, the non-LTE inversion yields same-sign line-of-sight
field that corresponds to the dominant polarity in either umbra of the
$\delta$-spot.
Similarly, this likely explains the discrepancy in azimuth direction for those
pixels.
Finally, the previously noted MERLIN saturation limit of 5\,kG means that the
strong line-of-sight and horizontal field that were inferred in both the LTE and
non-LTE inversions (see Figs.~\ref{fig:fe_fields} and \ref{fig:scatter}) cannot be
reproduced with the Milne-Eddington approach, where the line-of-sight field
ranges from $-$3.6\,kG to saturation at +5\,kG, also reaching saturation for the
horizontal field.

\paragraph{Chromospheric field}
\label{sec:results_wfa}
As the non-LTE inversions yield a depth-stratified atmosphere, we investigate
the response of the \CaII\ Stokes profiles to magnetic field changes and find
that this response peaks somewhere between $\ltau=-2.7$ and $-$4.3 depending on
the pixel in question.
We therefore decide to average the magnetic field components over seven depth
points centered at $\ltau=-3.5$, effectively covering $\ltau=[-2.9,-4.1]$.

\rev{The chromospheric line-of-sight field maps from the weak-field
approximation (Fig.~\ref{fig:mewfa_stic}a) and non-LTE inversions
(Fig.~\ref{fig:mewfa_stic}b) are largely the same, both in the distribution of
opposite polarities and in the strengths that they reach.
The evident exception is a band of positive polarity around \XYis{500}{-250} in
the non-LTE results, where the solution may have converged to a local minimum
and failed to fit all four Stokes parameters as well as outside this band.
In addition, the non-LTE inversion exhibits stronger field both below and
above the `head' of the inverse-S polarity inversion line, compared to the WFA
map.
The latter is naturally smoother given the spatial regularisation that couples
the solution from neighbouring pixels.}

Considering then the \rev{horizontal} field (green arrows) from the
spatially-regularised weak-field approximation
(Fig.~\ref{fig:mewfa_stic}\rev{a}) and non-LTE-inferred results
(\rev{Fig.~\ref{fig:mewfa_stic}b}), \rev{and comparing with that in the
photosphere}, 
we find that in both cases
for most of the positive polarity (right of $X=510$\arcsec) and part of the
negative polarity (between $X=510$\arcsec\ and 515\arcsec) 
the photospheric and chromospheric field azimuth \rev{point in nearly the same
direction}.
The discrepancies are larger in the top left of the \rev{SST field-of-view
(black box in the lower panels)---unsurprisingly,
as the field is weaker there---}but also close to the polarity inversion line for
the ME--WFA comparison, while the non-LTE-inferred azimuths are in closer
agreement between photosphere and chromosphere.

Comparing the chromospheric field from the WFA and non-LTE inversions we
also see that in both the horizontal component has a similar orientation in the
vicinity of the polarity inversion line and in strong line-of-sight field areas
in general.
This similarity holds also for the three-dimensional magnetic field vector, as
visualised in Fig.~\ref{fig:mewfa_stic}c, mapping the angle
difference $\theta$ between the non-LTE-inferred and WFA-derived magnetic field
vectors.
This angle $\theta$ is obtained simply as
\begin{equation}
  \label{eq:theta}
  \theta(\vec{B}_{1},\vec{B}_{2}) = \arccos\Bigg(\frac{\vec{B}_{1} \cdot
  \vec{B}_{2}}{|\vec{B}_{1}| |\vec{B}_{2}|}\Bigg)
\end{equation}
where in this case
$\vec{B}_{1} = \vec{B}_{\rm{WFA}}$ and  $\vec{B}_{2} = \vec{B}_{\rm{STiC}}$.
Over the full field-of-view more than 75\% of the pixels have an angle of
55\deg\ or less, but this statistic is in part driven by the large angle
differences that are found in weak-field regions (top left corner) where
derivation of the azimuth is not as reliable. 
In the strong-field parts of the field-of-view $\theta < 15\deg$ in general,
with the exception of the band of large $\theta$ around \XYis{500}{-250}, where
the non-LTE inversion (erroneously) returns a positive \Blos\ value.

Figure~\ref{fig:ca_scatter} quantifies the degree of similarity field further
and presents two-dimensional histograms for the line-of-sight and horizontal
field compontens, as well as the distributions of the angle $\theta$ between the
two three-dimensional magnetic field vectors and the field inclination as
derived with either method. 
In all three panels results are shown for those pixels for which the total WFA
magnetic field strength ($|\vec{B}_{\rm{WFA}}|$) exceeds 2\,kG (white contours
in Fig.~\ref{fig:mewfa_stic}c).
From the left-hand and middle panels we see that the WFA line-of-sight magnetic
field strength is strongly correlated with that from the non-LTE inversions, but
that the horizontal field strength presents a wide scatter cloud.
About 45\% of the selected pixels have a stronger line-of-sight component
in the non-LTE inversions than in the WFA, but slightly more than half have a
higher horizontal field (53\%) and total field (54\%) strength when inferred
with STiC.
The median non-LTE-to-WFA fractional difference is only a few percent for either
component: 1.2\% for \Blos, 3.0\% for \Bhor\ and 2.0\% for \Btot.
In addition, a few features stand out.
First, the line-of-sight panel indicates a higher density on the diagonal at
roughly $\pm$2\,kG, but these disappear when taking all pixels into account and
they are therefore a visualisation artefact from excluding pixels with
$|\vec{B}_{\rm{WFA}}| < 2$\,kG.
Second, the roughly horizontal scatter around $\BlosWFA=-2$\,kG and extending
over all positive \BlosSTiC\ values is due to pixels that have erroneously been
inferred with positive line-of-sight field strengths in the latter
(cf.~\rev{Fig.~\ref{fig:mewfa_stic}b and} the
aforementioned band of large $\theta$ around \XYis{500}{-250} in
Fig.~\ref{fig:mewfa_stic}c).

The right-hand panel shows that the distribution of angles $\theta$ between the
three-dimensional magnetic field vectors (solid blue line) is skewed to smaller
angles. 
The distribution has a median of 8\deg\ and over 91\% of the pixels have an angle
between the WFA and non-LTE-inferred magnetic field vectors of 25\deg\ or less.
Only 4\% of the pixels has an angle larger than 50\deg\ (\ie\ outside the
panel's range). 
The absolute difference between the WFA- and non-LTE-inferred field inclinations
($|\gamma_{\rm{WFA}}-\gamma_{\rm{STiC}}|$, dashed red line) is equally skewed,
with a median of 5\deg\ and over 90\% of the pixels with an inclination
difference of less than 15\deg.
Hence, although the horizontal field strengths are not as tightly correlated, the
per-pixel magnetic field orientiation is very similar between WFA and non-LTE
inversion.

Finally, Fig.~\ref{fig:profs_ca} presents a selection of \CaIR\ profile fits for
the coloured markers in Fig.~\ref{fig:mewfa_stic}b.
The first column samples a pixel in the positive-polarity umbra, while the other
four are of pixels in the \CaII\ flare ribbons, where the last two are for the
same red and blue pixels as in Fig.~\ref{fig:profs}.
The complex umbral profile (cyan) is well-fitted in all four Stokes components,
except for a spurious local maximum in Stokes $Q$.
Overall, the fits to Stokes $I$ and $V$ follow the observations closely
regardless of the intensities and profile shapes, while Stokes $Q$ and $U$
sometimes prove more difficult (\eg\ purple and blue samplings, in particular
the blue Stokes $U$), despite the high signal-to-noise ratio.
\rev{Systematic errors (\eg\ calibration errors due to remaining fringes or
variable seeing) in combination with a typically weaker signal compared to
Stokes $V$ are likely culprits for such occasionally poorer fits to Stokes $Q$
and/or $U$.}
The stronger horizontal chromospheric fields are in general only inferred when
Stokes $Q$ and $U$ are both well-recovered (\eg\ orange and red samplings). 
In some cases (\eg\ blue) the strong line-of-sight field leads to visibly
Zeeman-splitted Stokes $V$ lobes.

\paragraph{Magnetic field strengths}
While most of the field-of-view is inferred with relatively common field
strengths of up 2--3\,kG in both photosphere and slightly lower in the
chromosphere (cf.~the colour bars in Figs.~\ref{fig:fe_fields} and
\ref{fig:mewfa_stic}), certain pixels are inferred with values well in excess of
that.
These are typically found for flaring profiles in \FeI\ and \CaII\ and some
examples are given in Fig.~\ref{fig:profs} (last three columns) where $|\Btot|
\simeq 5.5-6$\,kG in the photosphere.
However, the contours in the middle panels of
Fig.~\ref{fig:mewfa_stic}---outlining places where the photospheric field
exceeds 4.5\,kG and the chromospheric field 3\,kG in \Blos\ (cyan) and the same
thresholds in \Bhor\ (green)---show that these are typically not isolated
pixels.
Moreover, while these are somewhat arbitrary thresholds, changing their values
does not change that this strong field is generally clustered in coherent
patches that persist from photosphere to chromosphere. 

\section{Comparison with numerical models}
\label{sec:results_inv_model}

We compare our magnetic field inference results with those from two 
numerical simulations, namely the magnetofrictional (MF) model from 
\citetads{2019A&A...628A.114P} 
(see also details in 
\citeads{2019SoPh..294...41P}) 
that was driven time-dependently with the inferred electric field
\citepads{2017SoPh..292..191L} 
from HMI vector magnetic field data,
and the magnetohydrodynamic model by
\citetads{2018ApJ...867...83I} 
that was initialised on a NLFFF extrapolation of the HMI photospheric field at
08:36\,UT.
For the former we consider the model snapshot at 09:24\,UT, which is the one
closest in time to the X2.2 flare peak, while for the latter we take the model
snapshot at $t = 0.28$\,h, equivalent to 08:52:48\,UT (just four minutes
before the X2.2 flare).
\rev{This choice is constrained by the cadence of the respective simulations and
our aim to compare with the observationally inferred field.
We furthermore note that selecting a different snapshot from the
magnetofrictional model would not significantly alter our comparison or
conclusions, given the temporal smoothing applied in its pre-processing (see
discussion in Section~\ref{sec:discussion_models}).}

For both we take the photospheric field from the $z=0$\,Mm height, while we
consider the average \CaIR\ formation height in our field-of-view to be between
1--2\,Mm (cf.~\eg\
\citeads{2019A&A...631A..33B}) 
and select
the \rev{only} height index that falls in that range in both simulations, resulting in
$z=1.75$\,Mm for the MF model and $z=1.44$\,Mm for the MHD
model.
\rev{This choice is again a limitation imposed by the model properties, but we
note that taking the chromospheric field as coming from one height index higher
does not significantly change the presented maps and that our choice also
minimises the height difference between the respective model slices.
Furthermore, the MHD model is not data-driven and does not aim to exactly
reproduce the temporal evolution of the observed events.
We therefore selected an already analysed snapshot (cf.
\citeads{2018ApJ...867...83I}) 
where the flux ropes still sit low in the atmosphere, as our observations also
suggest.}

\subsection{Inferred versus modelled magnetic field}
\begin{figure*}[bht]
  \centerline{\includegraphics[width=\textwidth]{\figspath/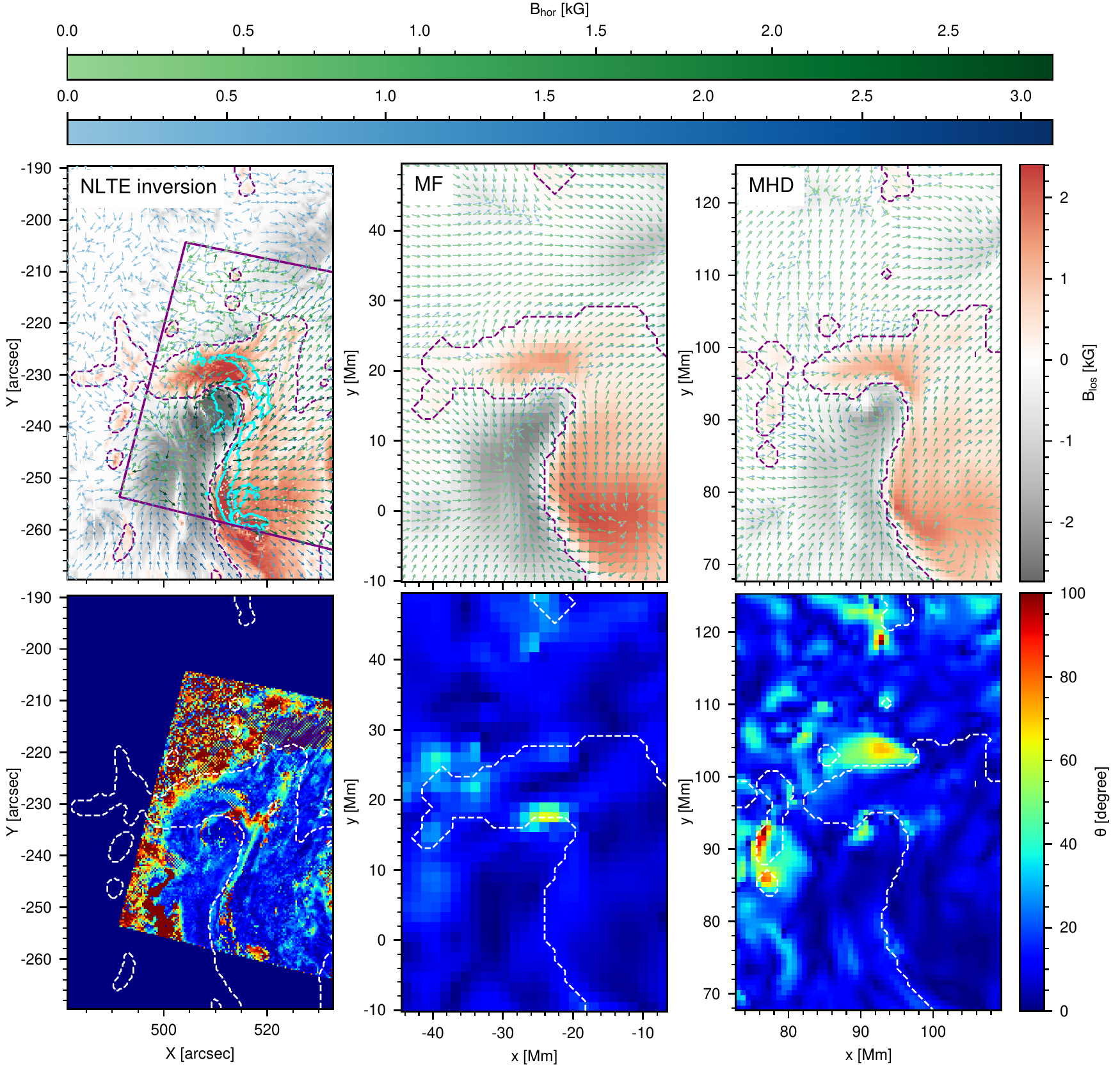}}
  \vspace{-2.5ex}
  \caption[]{\label{fig:inv_model} %
    Comparison of the non-LTE-inferred photospheric and chromospheric magnetic
    field ({\it left column\/}) with those from the magnetofrictional model
    ({\it middle column\/}) and the MHD model ({\it right column\/}).
    {\it Top row\/}: photospheric line-of-sight magnetic field with azimuth
    arrows for the photospheric field ({\it blue\/}) and chromospheric field
    ({\it green\/}), colour-scaled according to the corresponding colour bars.
    \rev{We encourage the reader to zoom in using a PDF viewer to appreciate the
    photosphere-to-chromosphere changes in magnetic field azimuth.}
    The inferred magnetic field has been taken at $\ltau=-1.1$ (from non-LTE
    \FeI) and $\ltau=[-2.9,-4.1]$ (from \CaIR) and the magnetic field map in the top
    left panel is identical to Fig.~\ref{fig:mewfa_stic}\rev{e, only including
    now also arrows for the chromospheric field azimuth (green)}.
    {\it Bottom row\/}: angle $\theta$ between the photospheric and
    chromospheric field vector of the inversions and models.
    The dashed lines (purple in the top row, white in the bottom row) indicate
    the photospheric line-of-sight polarity inversion lines, while the cyan
    contours in the top left panel outline the \CaII\ flare ribbons.
    }
\end{figure*}

Figure~\ref{fig:inv_model} presents a comparison of the non-LTE-inferred
photospheric and chromospheric field vector with that from the two numerical models.
The top row shows field vector maps, 
while the bottom row shows maps of the angle $\theta$ (cf.~Eq.~(\ref{eq:theta}))
between the three-dimensional photospheric and chromospheric field vectors for
the same three cut-outs.

The top middle and right-hand panels (\ie\ models) are \rev{relatively} similar
\rev{between each other}, yet differ considerably in several aspects from the
left-hand panel (\ie\ inferred field).
The overall distribution of positive and negative line-of-sight polarities is
similar between all three panels, but
where the non-LTE inference of the magnetic field yields absolute
line-of-sight field strengths in excess of 4\,kG in both photosphere and
chromosphere in certain places, the simulations do not reach beyond 2.4\,kG in
the photosphere and 1.3\,kG in the chromosphere anywhere.
Similarly for the horizontal field, the inferred values of up to 5\,kG in the
photosphere and 4.6\,kG in the chromosphere are 2--3 times larger than the
maxima in the simulations.
This is likely a combined effect of the pixel size difference between HMI
and \Hinode\ data, as well as the fact that the \Hinode\ data sample two
spectral lines with different Land\'e factors at high spectral resolution, 
leading to weaker field from HMI and consequently lower values in the models
that are based on that.

While more than half an hour apart, the two modelling results \rev{generally
exhibit the same line-of-sight field distribution reaching also similar
strengths} (both between about $\pm$2\,kG in the photosphere and $\pm$1.3\,kG in
the chromosphere\rev{, though slightly stronger in the MF model}).
\rev{The MF model also has more extended strong-field concentrations than the
MHD model.
Even larger differences are found} in the horizontal field strengths and
especially in the azimuth pattern.
Discrepancies in the latter are evident close to the polarity inversion line,
where the field in the MF model (middle panel) is oriented at a
larger angle to the polarity inversion line, while that of the MHD model (right
panel) follows the inverse-S shape more closely.
Also, the apparent source point of diverging positive field in the
MF model (at \xyisMm{-12}{0} in the middle panel) lies some
6--7\,Mm towards the North-West (\ie\ upper right) in the MHD model
(at \xyisMm{100}{74} in the right panel), while the `whirlpool'-like convergence
to negative line-of-sight field in the head of the inverse-S is similar in both,
though stronger in the MHD results.

In strong line-of-sight field regions the photospheric and chromospheric
azimuths are similar between the non-LTE inference and the models, whereas in
weaker-field areas the inferred azimuth appears more random. 
For all three panels, the photospheric and chromospheric azimuths are often
similar, but the inferred azimuths show typically the largest angles between the
photosphere and chromosphere---in particular in and close to the strong field
regions in the vicinity of the inverse-S shaped polarity inversion line.
Overall, and considering the strong-field part of the field-of-view in
particular, the inferred azimuths are best approached by those from the MHD
model.

\subsection{Photosphere-to-chromosphere field vector variation}
The smoothness in change of the field vector orientation from photosphere to chromosphere
in the models compared to the inversions is further emphasised in the bottom row
of Fig.~\ref{fig:inv_model}, showing maps of the angle
$\theta(\vec{B}_{\rm{phot}}, \vec{B}_{\rm{chro}})$
between the 3D photospheric and chromospheric magnetic field vectors.
For most of the strong-field regions of the field-of-view the angle between the
inferred field vectors is relatively small (below 50\deg), while strong
deviations are found outside the sunspot penumbra (above the upper polarity
inversion line, marked by the dashed white line).
Here the total field strength in both photosphere and chromosphere is small and
the azimuth consequently difficult to disambiguate, resulting in $\theta$
displaying a confetti-coloured randomness.
In the models most of the field-of-view has relatively shallow angles between
the photospheric and chromospheric field. 
Apart from the resolution and timing difference, both models tend to have the
enhancements in $\theta$ in similar places, although at often larger magnitude
in the MHD model (\eg\ the large patch of $\theta \simeq 50-100$\deg\ around
\xyisMm{94}{104}) and sometimes lacking a clear counterpart in the
MF model (\eg\ the band of $\theta \simeq 40-50$\deg\ around
\xyisMm{80}{75}).
What stands out in all three panels is the arched head of the inverse-S shape,
which shows angle differences between photosphere and chromosphere of at least
about 50--65\deg\ in the models and of 50--140\deg\ from the observations.
In the inferred field these enhanced angle differences are found along most of
the spine of the inverse-S (right of the dashed white line), where the
MF model exhibits only a haze of 20--30\deg\ and the MHD model
shows only a few localised enhancements of 30--50\deg.
Also, while the observations have the entire inverse-S head light up in
enhanced angles, the MF model is enhanced mostly in the eastern
half (\ie\ towards the left), while the MHD model has enhancements both at 
about \xyisMm{88}{92} and \xyisMm{97}{93}, with smaller angles in
between.

\section{Discussion}
\label{sec:discussion}
\subsection{Magnetic field approximations}
\label{sec:discussion_approx}
Both the Milne-Eddington and spatially-regularised weak-field approximation
offer a computationally inexpensive way of obtaining the magnetic field vector
from observations, compared to full-blown non-LTE inversions.
At the same time, they come with limitations, as we also see in this study.
In particular the \FeI\ Milne-Eddington inversion suffers from issues in
locations where the \FeI\ lines are in emission, recovering the wrong sign for
the line-of-sight polarity (cf.~Fig.~\ref{fig:fe_fields}) due to the linear
source function assumed in the model.
Unsurprisingly, these locations correspond well with those areas where white
light flare emission in the HMI pseudocontinuum (obtained in the vicinity of
\FeI\,\,6173\,\AA) has been reported in previous
studies (\eg\ Fig.~5 in 
\citeads{2019SoPh..294....4R} 
or Fig.~2 in 
\citeads{2018A&A...612A.101V}).  
Furthermore, comparison with (non-)LTE inversions reveals that the default 5\,kG
saturation limit imposed on the MERLIN Milne-Eddington line-of-sight and horizontal
field may be too stringent here---a limitation previously noted also in
non-flaring sunspots (\eg\ 
\citeads{2018ApJ...852L..16O})---
and this could play a role in general in flares that occur in similarly complex
configurations.

On the other hand, and despite the relatively coarse sampling of the \CaIR\
line, the spatially-regularised weak-field approximation does a good job at
recovering a magnetic field vector that is very similar to that obtained from
non-LTE inversions (cf.~Fig.~\ref{fig:ca_scatter}). 
Over 90\% of the pixels have the non-LTE and WFA inclinations within 15\deg\ 
and their 3D field vectors within 25\deg\ of each other.
The line-of-sight component in particular is close to the non-LTE-inferred value
in strong-field areas, while the transverse field exhibits a much larger
scatter.
This gets worse outside the sunspot, where the line-of-sight field is weak and
the field vector more horizontal, complicating retrieval of the transverse
component (as 
\citetads{2018ApJ...866...89C} 
already points out), even though the spatial constraint strongly mitigates the
adverse effects of noise
\citepads{2020arXiv200614487M}. 
More than half of the strong-field pixels have a stronger horizontal and
total field strength in the non-LTE inversions, while slightly less than half do
so for the line-of-sight field.

\subsection{Do we need non-LTE inversions of \FeI?}
\label{sec:discussion_nlte}
Several studies over the past 50 years have investigated the effects of assuming
LTE versus non-LTE in the formation of the \FeI\ lines (\eg\ 
\citeads{1972ApJ...176..809A}, 
\citeads{1973SoPh...32..283L}, 
\citeads{1982A&A...115..104R}, 
\citeads{1988ASSL..138..185R}, 
\citeads{2001ApJ...550..970S}, 
\citeads{2012A&A...547A..46H}, 
\citeyearads{2013A&A...558A..20H},  
\citeyearads{2015A&A...582A.101H}).  
Of particular interest for the present study are the effects on magnetic field
strength and inclination that are reported by 
\citetads{2020A&A...633A.157S}  
when inverting in LTE either LTE or non-LTE synthesised \FeI\ profiles. 
Hence, a natural course was to explore such effects for the data that we
analysed.

Qualitatively the differences in the inferred magnetic field components are minor
when considering our LTE and non-LTE \FeI\ inversion results
(cf.~Fig.~\ref{fig:fe_fields}). 
Although there are obvious per-pixel differences, a similar map of positive and
negative line-of-sight polarities is found for both and the locations of
stronger horizontal field coincide well between the two inversion approaches.
The latter is supported quantitatively by the typically tighter correlation
between LTE and non-LTE results for horizontal field strengths above
$\sim$4.5\,kG compared to those below (Fig.~\ref{fig:scatter}, second
row).
The largest differences are found in the inferred temperatures
(Fig.~\ref{fig:scatter}, last row) and field azimuth, with typically higher
temperatures by 500--1500\,K in the non-LTE results and a spatially smoother
counter-clockwise azimuth pattern in the negative \Blos\ polarity (\ie\ the
`head' of the inverse-S) compared to the LTE inversions. 
One reason for this could be that the temperature increase required to fit
the \FeI\ emission lines is larger in non-LTE than in LTE and if the placement
of such temperature gradient is limited by the node description,
this could lead to a discrepancy in the magnetic field as a result of a
difference in the shape of the source function.

Nonetheless, as far as the magnetic field is concerned there is no strong
indication that would favour non-LTE over LTE inversions of \FeI, while for
temperatures the differences can be significant, in particular at optical depths
$\ltau=-1$ and higher up in the atmosphere.
Whether this is worth a factor $\sim$2 increase in computational cost to perform
non-LTE inversions will thus depend on the particular scientific objective.

\subsection{Photospheric and chromospheric magnetic field}
\label{sec:discussion_field}
\paragraph{Field strengths}
In certain parts of the field-of-view the (non-)LTE inversions with STiC yield
stronger field than the Milne-Eddington inversion and weak-field approximation,
with values that are on the high end of (and for the chromospheric field much
larger than) what has generally been reported, even for sunspots.
For instance, the survey study by
\citetads{2006SoPh..239...41L}, 
analysing nearly 90 years worth of sunspot observations, emphasises this by
finding a mere 0.2\% of the sunspot group sample containing sunspots with
photospheric field strengths in excess 4\,kG, only one case of which at 6.1\,kG.
At the same time, several recent studies have reported strong fields in
sunspots.
\citetads{2018ApJ...852L..16O}, 
using a Milne-Eddington inversion, find field strengths of over 5--6\,kG between
two opposite-polarity umbrae, with the 6302.5\,\AA\ Stokes $I$ profile
exhibiting clear Zeeman splitting.
Field strengths of over 7\,kG have been inferred by
\citetads{2013A&A...557A..24V} 
and
\citetads{2019A&A...631A..99S} 
from \FeI\,\,6302\,\AA\ observations of sunspot penumbrae at locations that are
associated with strong downflows and where the consequent evacuation may explain
the large field strengths as probing sub-$\ltau=0$ heights.
\citetads{2020arXiv200312078C}, 
employing a similar inversion approach on a sunspot lightbridge, 
find field in excess of 5\,kG at all inversion nodes and even up to 8.25\,kG at
$\ltau=0$.

In flares, most photospheric field strengths have typically been derived from
SDO/HMI data.
For instance, 
\citetads{2012ApJ...748...77S} 
and
\citetads{2012ApJ...745L..17W} 
investigated the same X2.2 flare and reported \Bhor\ and \Blos\ with values of
1.5\,kG to over 2\,kG, while
\citetads{2016ApJ...828....4S} 
studied an M1.0 flare with underlying absolute line-of-sight field strengths of
up to 3\,kG. 
A C4.1 flare analysed by
\citetads{2016ApJ...819..157G} 
occurred in a $\delta$-spot with line-of-sight field up to about 1.5--2.0\,kG,
derived from both HMI and SST/CRISP \FeI\,\,6301.5\,\AA\ data.
Similar values have been reported from ground-based observations, \eg\  
based on LTE inversions of \SiI\,\,10827\,\AA\ 
\citetads{2015ApJ...799L..25K} 
found total field strengths of order 1--2\,kG in an M3.2 flare,
while 
\citetads{2017A&A...602A..60G} 
performed LTE inversion of \FeI\,\,10783\,\AA\ and \SiI\,\,10786\,\AA, yielding
\Blos\ and \Bhor\ of the order of 1.0--1.5\,kG
during an M1.8 flare in a $\delta$-spot configuration.
\citetads{2018ApJ...869...21L} 
reported \Bhor\ of 0.2--1.0\,kG and |\Blos| of 1.5--2.5\,kG in an M6.5 flare
observed in \FeI\,\,15648\,\AA. 
On the other hand, chromospheric field inferences in flares or filament
eruptions are scarce and
typically report values of a 0.3--2\,kG and rarely more, \eg\
\citetads{2014A&A...561A..98S} 
with a few hundred Gauss in a activated filament during a flare observed in
\HeI\,\,10830\,\AA,
\citetads{2017ApJ...834...26K} 
and
\citetads{2018ApJ...860...10K} 
with $\sim$1.5\,kG from \CaIR\ observations of an X1 and M1.9 flare, respectively, 
\citetads{2019A&A...621A..35L} 
with $\sim$2.5\,kG from \HeIDthree\ observations of a C3.6 flare, or
\citetads{2020arXiv200610473K} 
with up to 60\,G line-of-sight and up to 250\,G horizontal field from
\HeI\,\,10830\,\AA\ observations of an erupting filament.

For the particular active region under scrutiny here,
\citetads{2018A&A...620A.183J} 
report field strengths of order 2.5\,kG from LTE inversions both during and
after the X9.3 flare that followed the X2.2 flare, while
\citetads{2018RNAAS...2....8W} 
find transverse photospheric field of over 5.5\,kG at the PIL from direct measurement
of the Zeeman splitting in \FeI\,15648\,\AA\ GST spectra, a few hours
after the X9.3 flare.
In addition, 
\citetads{2019ApJ...880L..29A} 
report exceptionally strong, kilogauss-order coronal magnetic fields about 5.5\,h
prior to the X2.2 flare based on a NLFFF reconstruction that is able to
reproduce the gyroresonant emission observed with the Nobeyama Radio Heliograph.
Noteworthy is that their results support the $\sim$5.5\,kG field strengths
reported by 
\citetads{2018RNAAS...2....8W} 
and indicate field strenghts of order 3.5--3.0\,kG at 1--3\,Mm heights.
In this context it is therefore perhaps not surprising that we find places where
the horizontal and line-of-sight field strengths reach order 5--6\,kG in the
photosphere and 3--4\,kG in the chromosphere.

Given the exceptionally strong field that our inversions recovered, we also
considered a potential degeneracy between field strength and micro-turbulence.
Fitting a turbulence-broadened profile might cause the inversion code to settle
on a high magnetic field with low micro-turbulence, while observations in the
ultraviolet have been found consistent with the presence of micro-turbulence in
the chromosphere during (the onset of) flares at sites of both chromospheric
evaporation and condensation
(\eg\
\citeads{2011ApJ...740...70M}, 
\citeads{2013ApJ...774..122H}, 
\citeads{2018SciA....4.2794J}, 
\citeads{2020arXiv200405075G}). 
However, where these strongest field values are inferred in the photosphere, the
Zeeman splitting is often evident even in Stokes $I$ (Fig.~\ref{fig:profs}) and
the individual components are narrow, arguing against non-resolved motions.
Moreover, these pixels are largely found in coherent patches that are co-located
with similarly coherent strong-field patches in the chromosphere
(Fig.~\ref{fig:mewfa_stic}\rev{b and e}) and we therefore trust the values inferred for both
photosphere and chromosphere.

\paragraph{Height-dependent field vector}
The non-LTE inversions of \FeI\ and \CaII\ provide photospheric and
chromospheric field vectors and allow, for the first time, to track from
observations the orientation of the magnetic field vector with height in an
X-class flare.
In particular the chromospheric field azimuth, which derives from Stokes $Q$ and
$U$ that are often plagued by noise in the chromosphere, is notoriously
challenging to obtain even in flares and consequently few have tried
\citepads{2019A&A...621A..35L}. 
In the case of AR 12673 we benefit from the strong signal in these Stokes
components as a result of the underlying sunspot, which enables us to
confidently infer and disambiguate the chromospheric azimuth
(Fig.~\ref{fig:mewfa_stic}).

The angle between the three-dimensional photospheric and chromospheric field
vectors is small for most of the strong-field part of the field-of-view, while
there is an evident enhancement of some 40\deg--140\deg\ tracing the inverse-S
polarity inversion line and coinciding remarkably well with the flare ribbon
emission in \CaIR\ (Fig.~\ref{fig:inv_model}, left column panels).
While the magnetic flux rope system that has been proposed in various studies
(\eg\
\citeads{2018ApJ...867L...5L}, 
\citeads{2018ApJ...867...83I}, 
\citeads{2019SoPh..294....4R}, 
\citeads{2019A&A...628A.114P}, 
\citeads{2020ApJ...894...29B}) 
cannot be identified entirely in the inferred maps, it is nevertheless
worth noting that the two patches of enhanced $\theta$ in the lower right panel
of Fig.~\ref{fig:inv_model} appear to coincide with the footpoints of flux ropes
FR1 and FR4 in 
\citetads{2018ApJ...867...83I}. 
In addition, the concentrations of strong photospheric and chromospheric field
(Fig.~\ref{fig:mewfa_stic}\rev{b and e}) seem to be located at---or in the
vicinity of---the flux rope footpoints F1 and F2 in
\citetads{2020ApJ...894...29B}. 
We therefore speculate that the observed angle enhancements in the inverse-S
head (lower left panel) and the strong-field concentrations in the head and
lower down along the spine may similarly be sampling flux rope footpoints and
that our inferred chromospheric field is thus able to pick up at least part of
the flux rope system.

\subsection{Discrepancies with the numerical models}
\label{sec:discussion_models}

The most striking difference between the numerical models and inversions is in
the amplitudes of both the line-of-sight and transverse magnetic field. 
This can be partly attributed to the difference in spatial and spectral resolution of the
HMI and \Hinode\ SOT/SP instruments, but the pre-processing for both
simulations plays an equally---if not more important---role.
The latter is also evident when comparing the original HMI field vector
with the field from both numerical models at $z=0$~Mm. 
\citetads{2018ApJ...867...83I} 
use the pre-processing procedure by \citetads{2006SoPh..233..215W}, 
which modifies the observed magnetic field vector to fit the assumption of a
force-free field.  
As a result, the transverse field can get significantly reduced, in this case by
up to a factor 2 in some places. 
For the magnetofrictional model \citetads{2019A&A...628A.114P} 
apply spatial and temporal smoothing to ensure numerical stability of the
simulations. 
In addition, the vector magnetic field maps are rebinned spatially in both
approaches, by a factor 2 and 4 in the MHD and MF model, respectively. 
Combined, these effects result in a factor 2--3 in field strength amplitude
difference between the models and the inversions. 
We note here that we did not take the instrumental point spread function into
account and that use of spatially-coupled inversions
\citepads{2012A&A...548A...5V,2015A&A...577A.140A} would further increase the
difference. 

Obtaining the magnetic field vector accurately is extremely
important in space weather modelling and prediction.
Underestimating the field strength at the source has a direct impact on the
estimate for the flux carried by a pre-eruptive flux rope and would result in a
wrong estimate of its location and diameter, which in turn would produce a
mismatch between the models and the field measured at Earth
\citepads{2018ApJ...856...75T}. 
It is a well-recognized problem that an underestimate of the magnetic flux in
observationally retrieved maps can lead to an underestimate of the
interplanetary magnetic flux 
\citepads{2017ApJ...848...70L}. 
This is especially true for the quiet Sun, where most of the flux remains
invisible to current instruments \citepads{2016A&A...594A.103D}.  
However, our results indicate that this may even be the case in active regions, 
despite that the field there is strong and fills the whole surface area covered
by a pixel.
   
The second noticeable difference between the models and inversions is the strong
enhancement of some 40\deg--140\deg\ in the angle between the three-dimensional
photospheric and chromospheric field along the inverse-S polarity inversion
line. 
This band is conspicuously absent in the MHD model snapshot of
\citetads{2018ApJ...867...83I} 
which is from approximately 10\,min before our inversions (Fig.~\ref{fig:inv_model},
right column panels). 
Pre-processing could again be the culprit of this discrepancy, but it is
conceivable that the increase in photosphere-to-chromosphere shear may be due
to the field reconfiguration during the flare. 
The latter fits with the tether-cutting reconnection scenario proposed by  
\citetads{2019ApJ...870...97Z} 
in their two-step reconnection process, given that the soft X-ray flux
lightcurve goes through a steep increase between 09:08--09:10\,UT, where our
\CaII\ snapshot lies in the dead middle of. 
On the other hand, in their NLFFF extrapolation the reconnection that precedes
and triggers the tether-cutting reconnection occurs in a null point outside the
magnetic flux rope system, which unfortunately falls also outside our SST
field-of-view.  
The increase in retrieved photosphere-to-chromosphere shear is 
also in agreement with the post-flare configuration proposed by
\citetads{2020ApJ...894...29B}. 

Finally, an examination of the original HMI maps indicates that the position of
the diverging positive polarity of the inverse-S did not significantly change
in the time span of 09:12--09:24\,UT. 
This suggests that the mismatch in footpoint location between observations and
the MF model is likely due to the spatial smearing applied to the original field
maps as part of the pre-processing for the latter.

\section{Conclusions}
\label{sec:conclusions}
We have presented a comparison of the inferred photospheric and chromospheric
magnetic fields during the confined X2.2 flare in NOAA AR 12673 on September 6,
2017, allowing for the first time to track the variation in magnetic field vector
orientation from the photosphere to the chromosphere in an X-class flare.
Our results suggest that in the flare LTE formation of \FeI\ is not a
bad assumption {\it per se\/}, but that non-LTE inversions may yield stronger
line-of-sight field in the lower layers (with median differences of up to
7--9\%, depending on whether the full field or only flaring pixels are
considered) and higher temperatures throughout (up to 25\% for flaring pixels at
$\ltau=-2.1$).
Also, the disambiguated non-LTE field azimuth presents a smoother map in the
strongest-field areas than the LTE results.
Without knowledge of the true solution, however, we cannot rule in favour of one
or the other and performing a similar investigation of a flare simulation would
thus be worthwile.

On the other hand, we see that for this case Milne-Eddington inversions are an
oversimplification that suffer in the presence of \FeI\ emission profiles,
resulting in erroneous line-of-sight polarities.
Allowing for depth-dependence is ultimately necessary for a proper inference of
the magnetic field vector in this (and likely most) flaring region(s).
In the chromosphere, the spatially-constrainted weak-field approximation offers
an excellent estimate of the non-LTE-inferred magnetic field vector, where the
field strengths may be underestimated by a only few percent compared to actual
inversions.

While the chromospheric field points in approximately the same direction as the
photospheric field over most of the umbrae, there is a marked band of enhanced
angles (40\deg--140\deg) in the three-dimensional photosphere-to-chromosphere
field vector that closely traces the inverse-S polarity inversion line.
It coincides almost entirely with the location of the flare ribbons observed in
\CaIR\ and is therefore likely due to the flare-induced field reconfiguration.
During the flare there are coherent patches of strong line-of-sight and
horizontal field that persist from the photosphere (at $>4.5$\,kG) to
the chromosphere (at $>3$\,kG) and we find some pixels with either field
component in excess of even 5\,kG in both photospheric and chromospheric field
maps.
Both these strong-field concentrations and the enhanced
photosphere-to-chromosphere shear in the inverse-S head are found in close
proximity to flux rope footpoints proposed from modelling and our inversions
thus confirm such flux rope system configuration. 
However, compared to the models, the amplitudes of the inferred field strengths
are larger by up to a factor 2--3 and the photosphere-to-chromosphere shear is
also stronger, more concentrated and finely structured.
Both pre-processing for and lower spatial resolution of the numerical models are
likely culprits of these discrepancies.

Hence, while full-blown (non-)LTE inversions remain necessary to obtain a
depth-stratified atmospheric model, the spatially-regularised WFA represents a
powerful tool to quickly obtain the chromospheric magnetic field vector with
over a large field-of-view. 
In turn, this can be used as an additional boundary constraint in (flaring)
active region numerical modelling or help in improving existing models.
Validating or assessing these models remains very difficult, as one usually only
has photospheric observations and the coronal EUV emission, which the
(non-thermodynamic) models do not directly provide. 
Additional constraints in the form of chromospheric field maps are
therefore valuable.

The discrepancies in field strength and smoothness between the models and
inversions further emphasise the need for higher spatial resolution in the
\rev{models} to better constrain pre-eruptive flux ropes, as underestimating the
magnetic flux therein will impact the accuracy of CME evolution modelling in
space weather applications.

\begin{acknowledgements}
%
GV is supported by a grant from the Swedish Civil Contingencies Agency (MSB).
SD acknowledges the Vinnova support through the grant MSCA 796805.
This research has received funding from the European Union’s Horizon 2020
research and innovation programme under grant agreement No 824135.
JL is supported by a grant from the Knut and Alice Wallenberg foundation
  (2016.0019).
JdlCR is supported by grants from the Swedish Research Council (2015-03994), the
  Swedish National Space Board (128/15) and the Swedish Civil Contingencies
  Agency (MSB). This project has received funding from the European Research
  Council (ERC) under the European Union's Horizon 2020 research and innovation
  programme (SUNMAG, grant agreement 759548).
AR acknowledges support from STFC under grant No. ST/P000304/1.
This project has received funding from the European Research Council (ERC) under
  the European Union’s Horizon 2020 research and innovation programme (SolMAG,
  grant agreement No 724391). This project has received funding from the Academy
  of Finland (FORESAIL, grant number 312390; SMASH, grant number 310445).
We are grateful to H.~N.~Smitha at the Max-Planck-Institut f\"ur
Sonnensystemforschung for providing the \FeI\ model atom that we used in our
inversions and to Emilia Kilpua for useful comments.
The Institute for Solar Physics is supported by a grant for research
infrastructures of national importance from the Swedish Research Council
(registration number 2017-00625).
The Swedish 1-m Solar Telescope is operated on the island of La Palma by the
Institute for Solar Physics of Stockholm University in the Spanish Observatorio
del Roque de los Muchachos of the Instituto de Astrof\'isica de Canarias. 
\Hinode\ is a Japanese mission developed and launched by ISAS/JAXA, with NAOJ as
domestic partner and NASA and STFC (UK) as international partners. 
It is operated by these agencies in co-operation with ESA and NSC (Norway).
The SDO/HMI data used are courtesy of NASA/SDO and HMI science team.
The inversions were performed on resources provided by the Swedish National
Infrastructure for Computing (SNIC) at the National Supercomputer Centre at
Linköping University.
We thank R.~Shine as \Hinode\ SOT CO for scheduling the SOT/SP observations
  used here.
We made much use of NASA's Astrophysics Data System Bibliographic Services.
Last but not least, we acknowledge the community effort to develop open-source
  packages used in this work: 
  {\tt{numpy}} (\citeads{oliphant2006guide}; \url{numpy.org}), 
  {\tt{matplotlib}} (\citeads{Hunter:2007}; \url{matplotlib.org}), 
  {\tt{scipy}} (\citeads{2019arXiv190710121V}; \url{scipy.org}),
  {\tt{astropy}} (\citeads{astropy:2013}, \citeads{astropy:2018}; \url{astropy.org}),
  {\tt{sunpy}} (\citeads{sunpycommunity2020}; \url{sunpy.org}).

\end{acknowledgements}

\bibliographystyle{aa}

\bibliography{msb_flareinv}

\end{document}